\documentclass[11pt,leqno]{article}

\usepackage[margin=1.1in]{geometry}
\usepackage{amsmath,amssymb,amsthm}
\usepackage{graphicx}
\usepackage{float}
\usepackage{placeins}
\usepackage{xcolor}
\usepackage[authoryear,round]{natbib}

\usepackage{hyperref}
\hypersetup{
  pdftitle={Conformal prediction regions for dependent censored lifetimes},
  pdfauthor={Martin Bladt},
  colorlinks=true,
  citecolor=blue,     
  linkcolor=blue,     
  urlcolor=blue,      
  filecolor=blue,
}
\newtheorem{theorem}{Theorem}
\newtheorem{proposition}{Proposition}
\newtheorem{lemma}{Lemma}
\theoremstyle{definition}
\newtheorem{example}{Example}
\newtheorem{definition}{Definition}
\newtheorem{assumption}{Assumption}
\theoremstyle{plain}

\floatstyle{plaintop}
\newfloat{algorithm}{tbp}{loa}
\restylefloat{algorithm}
\floatname{algorithm}{Algorithm}

\newenvironment{assumptionlist}{%
  \begin{enumerate}%
  \setlength{\itemsep}{0.15em}%
}{\end{enumerate}}

\newcommand{\ind}{\mathbf 1}
\def\P{\mathbb P}
\def\E{\mathbb E}
\def\R{\mathbb R}
\def\tr{{\rm tr}}
\def\ca{{\rm ca}}
\newcommand{\Ztr}{\mathcal Z_{\tr}}
\newcommand{\Zca}{\mathcal Z_{\ca}}
\newcommand{\Ica}{\mathcal I_{\ca}}
\newcommand{\Icap}{\mathcal I_{\ca}'}
\newcommand{\indep}{\perp\!\!\!\perp}
\newcommand{\quantile}{\mathrm{Quantile}}

\newcommand{\rX}{\mathring X}
\newcommand{\rC}{\mathring C}
\newcommand{\rZ}{\mathring Z}

\makeatletter
\let\tableinput\@@input
\makeatother

\begin{document}

\title{Conformal prediction regions for multivariate and censored lifetimes}
\author{Martin Bladt\\[2pt]
\normalsize Department of Mathematical Sciences, University of Copenhagen\\
\normalsize Universitetsparken 5, 2100 K{\o}benhavn {\O}, Denmark\\
\normalsize \href{mailto:martinbladt@math.ku.dk}{\texttt{martinbladt@math.ku.dk}}}
\date{}
\maketitle

\begin{abstract}
Joint-life insurance and annuity benefits depend on two or more dependent lifetimes,
which portfolio data observe only up to a valuation date.  Given the covariates of a new
contract, such as the entry ages of a couple, we construct one lower bound per remaining
lifetime such that, with at least a prespecified probability, every life survives beyond its
bound.  The first step is to characterize the functions which allow for a useful projection from a multivariate to a univariate setting. The construction then relies on univariate conformal
prediction, which calibrates any fitted model on held-out data, so that this statement does
not rely on a correct mortality or dependence model.  Censoring is handled by calibrating
on the contracts observed for at least a fixed horizon, weighted by the inverse of their
probability of being observed that long.  We show that, when the lives can be censored at
different times, this is possible precisely for joint statements that every life survives beyond its
own bound, and that a common censoring time also permits bounds on the second-death
time. If censoring is independent of the lifetimes given the covariates, coverage holds
for every sample size when these probabilities are known, and approximately in large
samples when they or the conditional lifetime quantiles are estimated consistently.  As an application, we analyze a
Canadian joint-life annuity portfolio.
\end{abstract}

\noindent\textbf{Keywords:} conformal prediction; dependent lifetimes; joint-life
insurance; multivariate censoring; survival copula; valuation date.

\noindent\textbf{MSC:} 62G15; 62N01; 62H05.


\section{Introduction}

Joint-life insurance benefits depend on two lifetimes.  A first-to-die policy pays at
the first death and a last-survivor annuity pays until the second death, so the valuation of
either requires statements about the pair of remaining lifetimes of the insured couple,
whose dependence through shared circumstances and bereavement is well documented
\citep{FreesCarriereValdez1996,DenuitEtAl2001,SpreeuwOwadally2013}.  Portfolio data
have a particular structure.  Each contract is followed from entry into the
portfolio up to a valuation date, so that every lifetime is subject to right censoring at a time
that is known for every contract, and both lives are often censored at the same date.  With right-censored data, the natural
questions we are interested in answering are those regarding lower bounds on the lifetimes, and this paper constructs such bounds. Namely, given
the covariates of a new contract, such as the entry ages of a couple, every life survives
beyond its bound with probability at least $1-\alpha$, for instance $90\%$.  This
probability averages over new contracts from the portfolio population and over the data
used to compute the bounds.  We
call this property coverage at level $1-\alpha$, see \eqref{criterion} below.  It holds whatever
the joint distribution of the lifetimes and whether or not the fitted lifetime model is
correct, provided censoring is independent of the lifetimes given the covariates and the
probability that a contract is observed beyond a chosen horizon, given its covariates, is
known. A poor model gives less informative bounds and coverage that varies more across
ages, but not a lower average coverage.

Parametric and semiparametric copula models for coupled lives as in
\citep{FreesCarriereValdez1996,Carriere2000,LucianoSpreeuwVigna2008} specify joint
distributions, and multistate models \citep{Norberg1989,JiHardyLi2011} describe the
mechanism of dependence, but the validity of the resulting predictions depends on the correct
specification of the model, in particular of the margins and dependence structure.  In contrast, conformal prediction
\citep{VovkGammermanShafer2005,LeiEtAl2018} provides coverage for every sample size
around an arbitrary fitted model. Moreover, weighted versions of conformal prediction allow for the held-out contracts to have a known, different
covariate distribution \citep{TibshiraniEtAl2019}.  For a single
right-censored lifetime with observed censoring times, \citet{CandesLeiRen2023} studied conformal prediction using a clever technique using capping at a fixed threshold, together with inverse-probability-of-censoring techniques. Later work then also allows for covariate-dependent horizons,
\citep{GuiEtAl2024}, doubly robust calibration under covariate shifts \citep{YangEtAl2024}
and resampling-based calibration \citep{QinEtAl2025}.  Multivariate conformal regions
have been studied for fully observed responses, through copulas, ellipsoids, learned
representations and optimal-transport ranks
\citep{MessoudiEtAl2021,MessoudiEtAl2022,FeldmanBatesRomano2023,ThurinNadjahiBoyer2025}, but
conformal prediction regions for vectors of right-censored lifetimes have, to our
knowledge, not been treated.

For several lives, we apply the univariate construction to a real-valued function $s(X)$ computed from
the vector $X$ of remaining lifetimes. In actuarial terms, $s(X)$ is the (univariate) failure time of a
status \citep[Ch.~9]{Bowers1997}, such as $\min(X^{(1)},X^{(2)})$ for the joint-life status
and $\max(X^{(1)},X^{(2)})$ for the last-survivor status.  If $C$ is the vector of censoring
times, the observed vector is the componentwise minimum $Z=X\wedge C$, and the construction
requires $s(X\wedge C)=s(X)\wedge s(C)$, so that $s(Z)$ is nothing else than the new observation $s(X)$ censored at the the new censoring time $s(C)$.  With a common valuation date, the requirement  holds for both examples of $s$, but when the lives
are censored at different times, it fails for the last-survivor status.  We actually characterize this and show that it holds for any censoring
vectors exactly when $s(x)=\min_j\psi_j(x^{(j)})$ for nondecreasing functions $\psi_j$.  We call $\psi=(\psi_1,\ldots,\psi_d)$ a path, and it fixes
how the bounds of the lives move together, and the identity path gives all lives the same
bound, which retrieves the first-death time case.  However, with a common valuation date, every
order statistic can be used, and, as mentioned, in particular the second-death time
(see Proposition~\ref{prop:common}).  We use the same capping approach as in the univariate case for the univariate outcome $s(Z)$, and this gives coverage for every sample size, as long as certain retention probabilities are known. Else, coverage is approximate in large samples if
either these retention probabilities or the conditional quantiles of $s(X)$ are estimated
consistently (this is called double-robustness).  We also quantify the loss incurred by naively treating censored lifetimes as deaths, and compare with aggregation of univariate confidence regions using a  Bonferroni approach.  Finally, we study a Canadian joint-life insurance dataset.

\section{Observation model and compatible maps}\label{sec:setup}

This section fixes the structure of portfolio data observed at a valuation date and
identifies the maps for which censoring of the individual lives reduces to censoring
of a single duration.

\subsection{Setting}\label{sec:problem}

Fix the number of lives per contract, $d\geq2$, and the miscoverage level
$\alpha\in(0,1)$.  Let $W\in\R^p$ be the
covariates of an observation, that is, of a contract, and let
$X=(X^{(1)},\ldots,X^{(d)})$ and $C=(C^{(1)},\ldots,C^{(d)})$ be its vectors of
remaining lifetimes and of censoring times, measured from entry, both in $\R^d$.  For i.i.d.\ observations indexed by
$\mathcal I$, the data are
\begin{align}
 \mathcal Z=(W_i,C_i,Z_i)_{i\in\mathcal I},\qquad Z_i=X_i\wedge C_i,
 \label{data}
\end{align}
where the minimum is componentwise.  Dependence within $X$ and within $C$ is
unrestricted; in particular, all coordinates of $C$ may coincide, as they do when all
lives of a contract are censored at the valuation date.  We assume that $C$ is observed
for every observation, as is the case for portfolio data extracted at a valuation date or
for contract end dates fixed at issue \citep{MacdonaldRichardsCurrie2018}; this is
Type-I censoring.  Lapse and surrender are not of this type, since they are not observed
after a death.

We use the following two censoring conditions, the second one implying the first one, and invoke the stronger one only where
stated.
\begin{assumption}[conditional independence]\label{as:cic}
$X\indep C\mid W$.
\end{assumption}
\begin{assumption}[complete independence]\label{as:coic}
$(X,W)\indep C$.
\end{assumption}

A region rule assigns a set $\widehat R(\mathcal Z;w)\subseteq\R^d$ to the data and a
value of the covariates, with
$(\mathcal Z,w,x)\mapsto\ind\{x\in\widehat R(\mathcal Z;w)\}$ jointly measurable.  We
suppress $\mathcal Z$ and write $\widehat R(w)$.  The rule is \emph{valid} at level
$1-\alpha$, or has coverage $1-\alpha$, for a joint distribution of $(W,C,X)$ if
\begin{align}
 \P\bigl(X\in\widehat R(W)\bigr)\geq1-\alpha ,
 \label{criterion}
\end{align}
the probability being over $\mathcal Z$ and an independent copy $(W,C,X)$, which generates the new
observation.

\subsection{Censoring-compatible maps}\label{sec:shape}

To study the multivariate case using univariate techniques, we require a reduction map $s$, for example the failure time of a status, which gives the reduced
lifetime $s(X)$. However, not all $s$ are easy to analyze mathematically, and we first have to study this compatibility.

For $x,c\in\R^d$, let $x\wedge c$ denote the componentwise minimum, let $\leq$ denote the
componentwise order, and let $\bar\R=[-\infty,\infty]$.  A set $A$ in $\R$ or in $\R^d$
is an \emph{up-set} if $x\in A$ and $x\leq y$ imply $y\in A$; the up-sets of $\R$ are
$\emptyset$, $\R$ and the intervals $[a,\infty)$ and $(a,\infty)$.

\begin{definition}\label{def:compat}
A map $s:\R^d\to\bar\R$ is \emph{censoring-compatible} if
\begin{align}
 s(x\wedge c)=s(x)\wedge s(c)\qquad\text{for all }x,c\in\R^d .
 \label{compat}
\end{align}
\end{definition}

Identity \eqref{compat} makes $s(Z)$ a right-censored observation of $s(X)$ with
observed censoring variable $s(C)$.

\begin{theorem}\label{thm:shape}
For $s:\R^d\to\bar\R$ the following are equivalent.
\begin{itemize}
\item[\rm(i)] $s$ is censoring-compatible.
\item[\rm(ii)] There exist nondecreasing $\psi_1,\ldots,\psi_d:\R\to\bar\R$ with
 \begin{align}
  s(x)=\min_{1\leq j\leq d}\psi_j\bigl(x^{(j)}\bigr),\qquad x\in\R^d .
  \label{minform}
 \end{align}
\item[\rm(iii)] For every $v\in\R$ the upper level set $\{s\geq v\}$ is a product
 $\prod_{j=1}^dI_j(v)$ of one-dimensional up-sets $I_j(v)\subseteq\R$.
\item[\rm(iv)] $s$ is nondecreasing and satisfies the \emph{retention identity}
 \begin{align}
  s(x\wedge c)\wedge c_0=s(x)\wedge c_0
  \qquad\text{for all }c_0\in\R\text{ and all }x,c\in\R^d\text{ with }s(c)\geq c_0 .
  \label{retention}
 \end{align}
\end{itemize}
\end{theorem}
In words, (ii) identifies a joint-life status in which life $j$ runs on its own clock
$\psi_j$, and (iv) says that $s(X)\wedge c_0$ is observed whenever $s(C)\geq c_0$, which is
what the calibration rule give below use. However, when all lives of a contract are censored at the same date, less is needed as in shown in the next result.

\begin{proposition}[common valuation date]\label{prop:common}
Suppose $C=A\mathbf 1$ almost surely for a real random variable $A$, where
$\mathbf 1=(1,\ldots,1)$, and let
$s:\R^d\to\bar\R$ be measurable and nondecreasing with
\begin{align}
 s(x\wedge t\mathbf 1)=s(x)\wedge s(t\mathbf 1),\qquad x\in\R^d,\ t\in\R .
 \label{diagcompat}
\end{align}
Then $s(Z)=s(X)\wedge s(C)$ almost surely.  For every $c_0\in\R$, on
$\{s(C)\geq c_0\}$ one also has $s(Z)\wedge c_0=s(X)\wedge c_0$ almost surely.
\end{proposition}

Every order statistic $s(x)=x_{(k)}$, the time of the $k$th death, $k=1,\ldots,d$,
satisfies \eqref{diagcompat}, since $(x\wedge t\mathbf 1)_{(k)}=x_{(k)}\wedge t$.
Membership in $\{x_{(k)}\geq v\}$ means that at least $d-k+1$ lives survive to $v$;
for $k>1$ it is a union of upper orthants, sets of the form $\prod_j[a_j,\infty)$, rather than a single orthant as in
Theorem~\ref{thm:shape}.

A further two-life example can combine requirements on both death times, as follows.  For fixed
$a\geq0$, let $s_a(x)=\min\{x_{(2)},x_{(1)}+a\}$, the time at which payments stop under
a joint-and-survivor annuity whose survivor benefit lasts at most $a$ years.  Writing
$m=x_{(1)}$ and $M=x_{(2)}$ gives
$s_a(x\wedge t\mathbf 1)=\min\{M,t,m+a,t+a\}=s_a(x)\wedge t$,
and $s_a(t\mathbf 1)=t$, so \eqref{diagcompat} holds.  For $v\in\R$, the region
$\{s_a\geq v\}=([v,\infty)\times[v-a,\infty))\cup([v-a,\infty)\times[v,\infty))$ requires
both lives to survive to $v-a$ and at least one to $v$.  For $a>0$ the region is
non-convex; at $a=0$ the reduced lifetime is the first-death time.

Identity \eqref{diagcompat} requires a common clock.  Transforming all lifetimes by the
same nondecreasing function preserves it for every order statistic, but different
transformations for different lives, or measuring lifetimes as attained ages, make the
censoring times of the lives differ, and then only the minimum-type maps of
Theorem~\ref{thm:shape} remain available.

We call a $d$-tuple $\psi=(\psi_1,\ldots,\psi_d)$ of nondecreasing maps a \emph{path};
the \emph{identity path} has $\psi_j(t)=t$ for every $j$, and its corresponding reduced lifetime is the first-death
time $\min_jx^{(j)}$.  For a given path, we write $s_\psi$ for the reduction map \eqref{minform} and
\begin{align}
 R_\psi(v)=\bigl\{x\in\R^d:\ \psi_j(x^{(j)})\geq v,\ j=1,\ldots,d\bigr\}=\{s_\psi\geq v\}
 \label{orthant}
\end{align}
for $v\in\bar\R$, so in particular $R_\psi(-\infty)=\R^d$.  When each $\psi_j$ is
continuous and strictly increasing and $v$ lies in their common range,
$R_\psi(v)=\prod_j[\psi_j^{-1}(v),\infty)$ is a closed upper orthant with \emph{corner}
$\psi^{-1}(v)=(\psi_1^{-1}(v),\ldots,\psi_d^{-1}(v))$, and $v\mapsto\psi^{-1}(v)$ is an
increasing curve in $\R^d$.  The corner holds the lower bounds of the individual lives;
since many corners give the same coverage, the path fixes how a higher bound for one life is
traded against a lower bound for another.  We write
\begin{align}
 \rX=s_\psi(X),\qquad \rC=s_\psi(C),\qquad \rZ=s_\psi(Z),
 \label{ring}
\end{align}
so that $\rZ=\rX\wedge\rC$ by Theorem~\ref{thm:shape}, and both $\rZ$ and $\rC$ are
observed.  For a fixed path and $v\in\bar\R$, the conditional orthant probabilities of
the lifetime and censoring vectors are
\begin{align}
 \bar F_\psi(v\mid w)=\P(\rX\geq v\mid W=w),\qquad
 \bar G_\psi(v\mid w)=\P(\rC\geq v\mid W=w),
 \label{FGcond}
\end{align}
with unconditional versions $\bar F_\psi(v)=\P(\rX\geq v)$ and
$\bar G_\psi(v)=\P(\rC\geq v)$.


\section{Conformal calibration}\label{sec:method}

For a fixed path fixed, the task is to choose a cutoff $v$ from censored data so that the region
$R_\psi(v)$ can have coverage $1-\alpha$. This section provides the details of how to apply conformal calibration. Since the multivariate vector has been reduced to a univariate problem, most of the techniques here follow the usual univariate route, and the interesting part lies in connecting the output of such procedure back to the multivariate problem.

\subsection{Retention and dealing with a possible covariate shift}\label{sec:reduction}
For a fixed path $\psi$ and horizon $c_0\in\R$, let $Y=\rX\wedge c_0$, the reduced
lifetime capped at $c_0$.  Only observations with $\rC\geq c_0$,
called retained, are used for calibration, and the resulting bound cannot exceed $c_0$.  The
next result gives the distribution of $(W,Y)$ among the retained observations.

\begin{theorem}\label{thm:reduction}
On $\{\rC\geq c_0\}$, one has $\rZ\wedge c_0=Y$.  Under Assumption~\ref{as:cic}, if
$\bar G_\psi(c_0)>0$, then
\begin{align}
 P_{(W,Y)\mid\rC\geq c_0}
 =P_{W\mid\rC\geq c_0}\times P_{Y\mid W}.
 \label{shift}
\end{align}
If also $\bar G_\psi(c_0\mid W)>0$ almost surely, then $P_W\ll P_{W\mid\rC\geq c_0}$
with
\begin{align}
 \frac{dP_W}{dP_{W\mid\rC\geq c_0}}(w)
 =\frac{\bar G_\psi(c_0)}{\bar G_\psi(c_0\mid w)}.
 \label{LR}
\end{align}
Under Assumption~\ref{as:coic}, this likelihood ratio is one.
\end{theorem}

Here $P_{W\mid\rC\geq c_0}\times P_{Y\mid W}$ denotes the distribution with marginal
$P_{W\mid\rC\geq c_0}$ and conditional distribution $P_{Y\mid W}$.  In other words,
among the retained observations the covariate distribution differs from that of a new
contract (covariate shift), whereas the conditional distribution of the capped outcome given the
covariates is unchanged.  The reduced variables satisfy $\rX\indep\rC\mid W$ under
Assumption~\ref{as:cic}, since they are separate functions of $X$ and $C$, and we also have
$\{X\in R_\psi(v)\}=\{\rX\geq v\}$.  Thus a lower prediction bound for $\rX$ gives a prediction region for $X$, and \eqref{LR} gives the weights needed to calibrate on the
retained sample.  The denominator in \eqref{LR}, which we call the retention
probability, is the regression function of the
observed indicator $\ind\{C\in R_\psi(c_0)\}$ on $W$, so that estimating the weights may be reduced to an ordinary
binary regression and involves no survival techniques.

\begin{example}[covariate-dependent retention]\label{ex:covariateretention}
Let $W\in\{0,1\}$, use the identity path, and suppose
$C=(A,\ldots,A)$, where $A\mid W=w\sim\operatorname{Exp}(\lambda_w)$ and
$X\indep A\mid W$.  For $c_0\geq0$,
\begin{align*}
 \bar G_\psi(c_0\mid w)=e^{-\lambda_wc_0},
\end{align*}
so the likelihood ratio in \eqref{LR} is proportional to $e^{\lambda_wc_0}$.  If
$\P(W=1)=p$, then
\begin{align*}
 \P(W=1\mid\rC\geq c_0)
 =\frac{p e^{-\lambda_1c_0}}
 {(1-p)e^{-\lambda_0c_0}+p e^{-\lambda_1c_0}}.
\end{align*}
Thus retention changes the covariate distribution when $\lambda_0\ne\lambda_1$, and
the likelihood ratio is designed to correct this shift.
\end{example}

\subsection{Conformal calibration for multivariate outcomes}\label{sec:algo}
In split-conformal calibration \citep{VovkGammermanShafer2005,LeiEtAl2018}, the data
are divided into two parts, a training fold, on which any model is fitted, and a calibration fold, on which a
nonconformity score measures how far each observed outcome falls below the fitted
prediction; the prediction set for a new observation consists of the outcome values whose
score does not exceed an empirical quantile of the calibration scores.  When the
calibration observations have a different covariate distribution than the new observation, as the
retained observations do here, the distribution shift is accounted for by weighting each calibration
score by the likelihood ratio of the two covariate distributions, see \citep{TibshiraniEtAl2019}.

Partition the sample into independent training and calibration folds, denoted by
$\Ztr$ and $\Zca$, of sizes $N=|\mathcal I_{\tr}|$ and $n=|\Ica|$.  Using the training
fold, we select a path $\hat\psi$, a finite horizon $c_0$, and a measurable nonconformity score
$V(w,y;\Ztr)$ that is nonincreasing in $y$, extended to $y=-\infty$ by
$V(w,-\infty;\Ztr)=\sup_{y\in\R}V(w,y;\Ztr)$ so that scores take values in $\bar\R$.
Choices that are measurable functions of $\Ztr$, that is, computed from the training
fold alone, are called \emph{training-measurable}.
The reduction map determines the family of prediction regions; the nonconformity
score is used to calibrate the prediction bound.
Large scores indicate outcomes below the fitted prediction.  For example, conformalised quantile regression
\citep{RomanoEtAl2019} uses $V(w,y;\Ztr)=\hat q(\alpha\mid w)-y$, where $\hat q$ is a
finite-valued estimate of the conditional $\alpha$-quantile of the capped reduced
outcome.

In the calibration construction and its asymptotic analysis, write
$\hat s=s_{\hat\psi}$ and suppress the dependence of the following quantities on
the training-selected path and horizon.  Conditional on $\Ztr$, define
\begin{align}
 \bar F(v\mid w)&=\bar F_{\hat\psi}(v\mid w)
 =\P(\hat s(X)\geq v\mid W=w,\Ztr),\nonumber\\
 \pi(w)&=\bar G_{\hat\psi}(c_0\mid w)
 =\P(\hat s(C)\geq c_0\mid W=w,\Ztr),
 \label{weights}
\end{align}
where $\pi$ is the retention probability and $\hat \pi$ is a training-fold estimate
of it.  Multiplying all weights by a
positive constant leaves the construction unchanged, so the numerator of \eqref{LR} need not be estimated.  For a probability measure $Q$ on $\bar\R$, let
$\quantile(1-\alpha;Q)=\inf\{z\in\bar\R:Q([-\infty,z])\geq1-\alpha\}$, and let
$\delta_z$ be the point mass at $z$.

\begin{algorithm}[tbp]
\caption{Conformal prediction regions under multivariate censoring.}
\smallskip\hrule\smallskip
\emph{Input:} data $\mathcal Z$; level $\alpha\in(0,1)$; covariates $w$ of the new
observation; training rules for the path, horizon, score and nonnegative finite weight
function $\omega$.
\begin{enumerate}
\item Split $\mathcal Z$ into $\Ztr$ and $\Zca$.  Using $\Ztr$ alone, select
 $\hat\psi$, $c_0$ and $V$; fix a weight function $\omega$, which may depend on $\Ztr$ and
 on the retention probabilities; put $\hat s=s_{\hat\psi}$.
\item Retain $\Icap=\{i\in\Ica:\hat s(C_i)\geq c_0\}$ and compute
 $Y_i=\hat s(Z_i)\wedge c_0$ and $V_i=V(W_i,Y_i;\Ztr)$ for $i\in\Icap$.
\item Put $D(w)=\sum_{i\in\Icap}\omega(W_i)+\omega(w)$.
 If $D(w)>0$, set
 \begin{align*}
  Q_w=\frac{\sum_{i\in\Icap}\omega(W_i)\delta_{V_i}
                +\omega(w)\delta_{+\infty}}{D(w)};
 \end{align*}
 otherwise set $Q_w=\delta_{+\infty}$.
\item Set $\eta(w)=\quantile(1-\alpha;Q_w)$ and the lower prediction bound
 \begin{align*}
  \widehat L(w)=\inf\{y\in\R:V(w,y;\Ztr)\leq\eta(w)\}\wedge c_0.
 \end{align*}
\end{enumerate}
\emph{Output:} $\widehat R(w)=R_{\hat\psi}(\widehat L(w))$.
\smallskip\hrule
\label{alg:main}
\end{algorithm}

Algorithm~\ref{alg:main} implements the multivariate reduction and then applies the
univariate construction of \citet{CandesLeiRen2023} to the reduced outcome.  The atom at
$+\infty$ replaces the unknown score of the new observation by its worst case; with equal
weights and $m$ retained observations it has mass $1/(m+1)$.  For the quantile-regression
score,
\begin{align}
 \widehat L(w)=\{\hat q(\alpha\mid w)-\eta(w)\}\wedge c_0.
 \label{CQRbound}
\end{align}
Other choices include a mean-regression score $\hat m(w)-y$ and a distribution-regression
score $\widehat{\bar F}(y\mid w)-(1-\alpha)$ with $\widehat{\bar F}(\cdot\mid w)$
nonincreasing \citep{ChernozhukovWuthrichZhu2021}.  The finite-sample result below permits any such score; we use quantile
regression for the asymptotic analysis.  For the orthant family, the prediction event is
$\{X\geq\hat\psi^{-1}(\widehat L(W))\}$, with the marginal coverage specified below.
Bounds for different $\alpha$ need not be ordered and should not be read as quantiles of a
predictive lifetime distribution.

Since $\widehat L(w)\leq c_0$, the reported region contains $R_{\hat\psi}(c_0)$.  When
$\widehat L(w)=c_0$, the output is $R_{\hat\psi}(c_0)$, the smallest region
available at horizon $c_0$, and has coordinatewise lower bounds
$\hat\psi^{-1}(c_0)$.  When $\widehat L(w)=-\infty$, the output is $\R^d$, which provides no
information and which we call \emph{uninformative}.  Increasing $c_0$ allows a
higher prediction bound but reduces retention.
Both the path and the horizon must be selected without using the calibration fold.

\subsection{Finite-sample marginal coverage for multivariate outcomes}\label{sec:validity}
The following theorem establishes finite-sample coverage for the reduction described in
the previous section.
\begin{theorem}\label{thm:validity}
Let $\widehat R$ be the output of Algorithm~\ref{alg:main}.
\begin{itemize}
\item[\rm(i)] Under Assumption~\ref{as:coic}, with $\omega\equiv1$,
 $\P(X\in\widehat R(W))\geq1-\alpha$.
\item[\rm(ii)] Under Assumption~\ref{as:cic}, if $\pi(W)>0$ almost surely for almost
 every training fold, then with $\omega(w)=1/\pi(w)$,
 $\P(X\in\widehat R(W))\geq1-\alpha$.
\end{itemize}
Both conclusions hold for every sample size and every training-measurable choice of
path, horizon and score satisfying the stated conditions.
\end{theorem}
Conditioning on the training fold fixes the path, so fitting it does not invalidate the
finite-sample coverage.  In contrast, substituting $\hat \pi$ for $\pi$ introduces
weight-estimation error, which is studied in Section~\ref{sec:dr}.  The condition
$\pi(W)>0$ in part (ii) is specific to the selected path and horizon, and does not
necessarily follow from the same condition for other choices.  If no observation is
retained, Algorithm~\ref{alg:main} returns $\R^d$.

Both conclusions also hold within any subpopulation $\{W\in B\}$ with $B$ being a set fixed on the training
fold and $\P(W\in B\mid\Ztr)>0$, whenever the algorithm is run on the calibration observations
with $W_i\in B$ and applied
to new observations with $W\in B$.  The proof conditions on $\Ztr$ and applies to the
conditional distribution given $W\in B$, under which Assumptions~\ref{as:cic} and~\ref{as:coic}
remain valid and the likelihood ratio \eqref{LR} changes only by a constant factor.
Consequently, calibrating separately within training-defined groups of covariates, such as age
bands, still gives coverage within each group, although at the expense of a smaller number of retained
observations.

\subsection{Asymptotic validity}\label{sec:dr}

For a nonincreasing function $\bar A$, define the generalised inverse
\begin{align}
 \bar A^{-1}(u)=\sup\{v:\bar A(v)\geq u\}.
 \label{ginv}
\end{align}
Set $Y=\hat s(X)\wedge c_0$, and define
\begin{align}
 q(w)=\bar F^{-1}(1-\alpha\mid w)\wedge c_0,\qquad
 e(w)=|\hat q(\alpha\mid w)-q(w)|.
 \label{qdef}
\end{align}
The quantities $\pi$, $\bar F$ and $q$ depend on the training fold through the selected
path and horizon.  The expectations below take averages over that fold and over an independent
$W$ from the target population, unless conditioning is explicitly stated.
\begin{assumption}[conditions on the weights and the quantile model]\label{as:nuisance}
\begin{assumptionlist}
\item\label{as:O} There are $\rho>0$ and $M<\infty$ such that $\pi,\hat \pi\in(0,1]$ and,
 for every $N$, almost surely,
 \begin{align*}
  \E[\pi(W)^{-1-\rho}\mid\Ztr]\leq M,\qquad
  \E[\hat \pi(W)^{-1-\rho}\mid\Ztr]\leq M.
 \end{align*}
\item\label{as:C} As $N\to\infty$,
 $\E[|\hat \pi(W)^{-1}-\pi(W)^{-1}|]\to0$.
\item\label{as:D} There are $b_2\geq b_1>0$ and $r>0$, independent of $N$ and
 the training fold, such that almost surely, for $P_W$-almost every $w$, $q(w)$ is
 finite, $q(w)+r<c_0$, and for $|t|\leq r$,
 \begin{align}
  b_1|t|\leq|\bar F(q(w)+t\mid w)-(1-\alpha)|\leq b_2|t|.
  \label{Dcond}
 \end{align}
\item\label{as:E} As $N\to\infty$, $\E[e(W)/\hat \pi(W)]\to0$.
\end{assumptionlist}
\end{assumption}

Assumption~\ref{as:O}, the inverse-moment condition, controls the tails of the true and
estimated inverse-retention weights.
Assumption~\ref{as:D} requires the conditional survival function to cross $1-\alpha$
with a slope bounded away from zero and infinity, strictly below the horizon $c_0$.  Assumption~\ref{as:E} measures quantile error with a weight that
penalises errors where estimated retention is low.

\begin{theorem}\label{thm:dr}
Suppose Assumptions~\ref{as:cic} and~\ref{as:O} hold, and use $\omega(w)=1/\hat\pi(w)$ in
Algorithm~\ref{alg:main}.
\begin{itemize}
\item[\rm(i)] If Assumption~\ref{as:C} holds, then, for any score permitted by
 Algorithm~\ref{alg:main}, that is, any measurable $V(w,y;\Ztr)$ nonincreasing in $y$,
 \begin{align*}
  \liminf_{N,n\to\infty}\P(X\in\widehat R(W))\geq1-\alpha.
 \end{align*}
\item[\rm(ii)] For the quantile-regression score, if Assumptions~\ref{as:D}
 and~\ref{as:E} hold, then, as $N,n\to\infty$,
 \begin{align}
  \widehat L(W)-q(W)&\overset{\P}\longrightarrow0,\label{qconsistency}\\
  \P(X\in\widehat R(W)\mid W,\Ztr,\Zca)
  &\overset{\P}\longrightarrow1-\alpha.\label{condcoverage}
 \end{align}
 In particular, marginal coverage converges to $1-\alpha$.
\end{itemize}
\end{theorem}

Together, the two parts give double robustness, since under Assumption~\ref{as:O}, coverage
holds asymptotically if either the retention probabilities (part (i)) or the conditional
quantiles (part (ii)) are estimated consistently. The
coverage given the covariates in \eqref{condcoverage} is asymptotic and depends on consistent
outcome-quantile estimation and the stated regularity conditions.

Under common-date censoring, the calibration and coverage results of this section
extend, under the same assumptions, to any training-selected reduction map $s$
satisfying \eqref{diagcompat}, with prediction regions $\{s\geq v\}$.
Proposition~\ref{prop:common} provides the identities
required by the construction and in the proofs.


\section{Dependence analysis and comparisons}\label{sec:price}
This section asks how the dependence between the censoring times of the lives affects
retention (Section~\ref{sec:retention}), how much is lost by treating censored lifetimes as
deaths (Section~\ref{sec:frontier}), and when the joint region improves on separate bounds
for the lives combined by Bonferroni's inequality (Section~\ref{sec:bonf}).  Because
these results assume a fixed path, they do not establish an ordering of finite-sample
fitted regions.  The marginal survival functions on the scale of the path, which we call
standardised, are denoted by
$\bar F_{\psi,j}(v)=\P(\psi_j(X^{(j)})\geq v)$ and
$\bar G_{\psi,j}(v)=\P(\psi_j(C^{(j)})\geq v)$, and the survival function of the
reduced censored outcome is $\bar H_\psi(v)=\P(\rZ\geq v)$.  Under
Assumption~\ref{as:coic}, $\rX$ and $\rC$ are independent, so
\begin{align}
 \bar H_\psi(v)=\bar F_\psi(v)\bar G_\psi(v).
 \label{ZisXC}
\end{align}
The target and observed-outcome cutoffs are, respectively,
\begin{align}
 v^F_\psi=\bar F_\psi^{-1}(1-\alpha),\qquad
 v^H_\psi=\bar H_\psi^{-1}(1-\alpha),
 \label{corners}
\end{align}
and $v^H_\psi\leq v^F_\psi$ since $\bar H_\psi\leq\bar F_\psi$.

\subsection{Dependence analysis using the censoring copula}\label{sec:retention}

Suppose the coordinates of $\psi$ are continuous and strictly increasing, the margins of
$C$ are continuous, and the horizons considered all lie within the range of every function $\psi_j$.
Let $\mathcal K_C$ be the survival copula of $C$.  Since the unconditional retention probability $\bar G_\psi(c_0)$ is the
orthant probability $\P(C\geq\psi^{-1}(c_0))$, Sklar's theorem for survival functions
and the Fr\'echet--Hoeffding bounds \citep[Ch.~2]{Nelsen2006} give
\begin{align}
 \bar G_\psi(c_0)
 &=\mathcal K_C\bigl(\bar G_{\psi,1}(c_0),\ldots,\bar G_{\psi,d}(c_0)\bigr),
 \label{piformula}\\
 \max\Bigl\{\sum_{j=1}^d\bar G_{\psi,j}(c_0)-(d-1),0\Bigr\}
 &\leq\bar G_\psi(c_0)\leq\min_j\bar G_{\psi,j}(c_0).
 \label{FH}
\end{align}
The upper bound in \eqref{FH} is attained when the coordinates of $C$ are comonotone
\citep{DhaeneEtAl2002}, in particular when all lives of a contract are censored at the
valuation date, whereas \eqref{piformula} reduces to the product of the margins when they
are independent.  This probability therefore depends on the path only through
the corner $\psi^{-1}(c_0)$, and on the censoring distribution through its margins at the
corner and its survival copula.  Conditional on a training-selected path and horizon,
the expected number of retained calibration observations is $n\bar G_{\hat\psi}(c_0)$.
To isolate the effect of dependence between the censoring times, the next result
assumes equal standardised margins.

\begin{proposition}\label{prop:retention}
Suppose the standardised censoring margins are equal, $\bar G_{\psi,j}=\bar G_0$ for
every $j$, and let $\pi_0\in(0,1)$.  The largest horizon with retention probability at
least $\pi_0$ is
\begin{align}
 \bar G_\psi^{-1}(\pi_0)=
 \begin{cases}
  \bar G_0^{-1}(\pi_0),&\text{if the coordinates of $C$ are comonotone},\\
  \bar G_0^{-1}(\pi_0^{1/d}),&\text{if they are independent}.
 \end{cases}
 \label{retentionfloor}
\end{align}
In particular, if $\bar G_0(c)=e^{-\gamma c}$ for $c\geq0$, the horizon is
$-\log\pi_0/\gamma$ for every $d$ under comonotone censoring and
$-\log\pi_0/(d\gamma)$ under independent censoring.
\end{proposition}

\begin{example}[common and coordinate-specific censoring]\label{ex:MO}
On the standardised scale, suppose $\psi_j(C^{(j)})=A\wedge B_j$, where $A$ is common to
all lives of the contract, such as the time to the valuation date, and $B_j$ is specific
to life $j$, such as the time to a contract end date fixed at issue, with
$A,B_1,\ldots,B_d$ independent exponential variables of rates $\lambda_0$ and $\lambda_1$
respectively; a zero rate denotes an infinite time, and $\lambda_0+\lambda_1>0$.  This is
the common-shock model of \citet{MarshallOlkin1967}, with
\begin{align}
 \bar G_{\psi,j}(c)=e^{-(\lambda_0+\lambda_1)c},\qquad
 \bar G_\psi(c)=e^{-(\lambda_0+d\lambda_1)c},\qquad c\geq0.
 \label{MOretention}
\end{align}
More generally, if $B_j$ has rate $\lambda_j$ and
$t=(t_1,\ldots,t_d)\in[0,\infty)^d$, then
\begin{align*}
 \P\bigl(\psi_j(C^{(j)})\geq t_j,\ j=1,\ldots,d\bigr)
 =\exp\Big\{-\lambda_0\max_jt_j-\sum_{j=1}^d\lambda_jt_j\Big\}.
\end{align*}
The term $\lambda_0\max_jt_j$ arises from the common censoring time $A$, and
$\sum_j\lambda_jt_j$ from the coordinate-specific times.  When $t_j=c$ and
$\lambda_j=\lambda_1$ for every $j$, the largest $c$ with retention probability at
least $\pi_0$ is $-\log\pi_0/(\lambda_0+d\lambda_1)$. 
\end{example}

\subsection{Comparison against naive calibration}\label{sec:frontier}
Since $Z\leq X$ and $R_\psi(v)$ is an up-set, coverage for the censored vector $Z$
implies coverage for $X$.  Ordinary conformal calibration applied to $Z$, without
retention or weights, which treats censored lifetimes as if death occurred at
censoring, is therefore valid, albeit conservative.  We now quantify this conservativeness.  In this subsection,
covariates are suppressed, and consequently $\widehat L$ is a function of the sample alone. We also assume that the cutoffs in \eqref{corners} are finite.

\begin{proposition}\label{prop:naive}
Let $\widehat L$ be a function of the sample alone.
\begin{itemize}
\item[\rm(i)] If $\P(Z\in R_\psi(\widehat L))\geq1-\alpha$, then, for every
 $\varepsilon>0$, with $p_\varepsilon=\bar H_\psi(v^H_\psi+\varepsilon)$, which is smaller
 than $1-\alpha$ by the definition of $v^H_\psi$,
 \begin{align*}
  \P(\widehat L>v^H_\psi+\varepsilon)\leq\frac{\alpha}{1-p_\varepsilon}.
 \end{align*}
 In particular, $\P(\widehat L\geq v)\leq\alpha$ if $v>\mathop{\rm ess\,sup}\rC$.
\item[\rm(ii)] Under Assumption~\ref{as:coic},
 \begin{align*}
  \bar F_\psi(v^H_\psi)\geq\frac{1-\alpha}{\bar G_\psi(v^H_\psi)},
 \end{align*}
 with equality if $\bar H_\psi$ is continuous at $v^H_\psi$.
\end{itemize}
\end{proposition}

Part (i) describes the sampling variability of a rule calibrated on $Z$.  For each
$\varepsilon>0$, it bounds the probability that the random cutoff $\widehat L$
exceeds the population cutoff $v^H_\psi+\varepsilon$.  Part (ii) instead concerns the fixed population cutoff
$v^H_\psi$.  If $\bar H_\psi$ is continuous at this point, then
\begin{align*}
 \P\bigl(X\in R_\psi(v^H_\psi)\bigr)
 =\bar F_\psi(v^H_\psi)
 =\frac{1-\alpha}{\bar G_\psi(v^H_\psi)}.
\end{align*}
Since $1-\alpha\leq\bar G_\psi(v^H_\psi)\leq1$, the corresponding miscoverage
$\alpha'=1-(1-\alpha)/\bar G_\psi(v^H_\psi)$ lies in $[0,\alpha]$.  It equals $\alpha$
when $\bar G_\psi(v^H_\psi)=1$, and approaches zero as $\bar G_\psi(v^H_\psi)$
approaches $1-\alpha$.

\begin{example}[exponential standardised margins]\label{ex:exponential}
Suppose Assumption~\ref{as:coic} holds and, for $v\geq0$, $\bar F_{\psi,j}(v)=e^{-v}$
and $\bar G_{\psi,j}(v)=e^{-\gamma v}$, with $\gamma>0$.  Suppose the survival copulas of
$X$ and of $C$ have diagonal sections $\mathcal K(u,\ldots,u)$ equal to $u^{\theta_X}$ and
$u^{\theta_C}$, respectively, the power-diagonal model.
The Fr\'echet--Hoeffding bounds imply $1\leq\theta_X,\theta_C\leq d$; comonotonicity
gives exponent $1$ and independence exponent $d$.  By Sklar's theorem,
$\bar F_\psi(v)=e^{-\theta_Xv}$ and $\bar G_\psi(v)=e^{-\theta_C\gamma v}$, so with
$\kappa(a)=-\log(1-a)$,
\begin{align}
 v^F_\psi=\frac{\kappa(\alpha)}{\theta_X},\qquad
 v^H_\psi=\frac{\kappa(\alpha)}{\theta_X+\theta_C\gamma},\qquad
 \frac{v^H_\psi}{v^F_\psi}=\frac{\theta_X}{\theta_X+\theta_C\gamma}.
 \label{ratio}
\end{align}
Accordingly, Proposition~\ref{prop:naive}\,(ii) gives
\begin{align*}
 \P\bigl(X\in R_\psi(v^H_\psi)\bigr)
 =(1-\alpha)^{\theta_X/(\theta_X+\theta_C\gamma)}>1-\alpha.
\end{align*}
For independent censoring coordinates, $\theta_C=d$, and the ratio is of order $1/d$ if
$\theta_X$ remains bounded as $d$ increases; if the lifetimes are also independent, the
ratio is $1/(1+\gamma)$.  Under a common valuation date, $\theta_C=1$, but the ratio can
still vary with $d$ through the dependence of the lifetimes.
\end{example}

\subsection{Comparison with Bonferroni bounds}\label{sec:bonf}
For each coordinate, a univariate lower bound with miscoverage at most $\alpha/d$ can
be combined into a joint region with coverage at least $1-\alpha$ by the union bound.
The $j$th bound can be calibrated using observations satisfying
$\psi_j(C^{(j)})\geq c_{0,j}$ for a coordinate-specific horizon $c_{0,j}$; joint
calibration with horizon $c_0$ instead retains only observations satisfying
$\psi_j(C^{(j)})\geq c_0$ for every $j$.  Quantities computed from the true
distribution are called oracle quantities.  To compare the resulting population regions,
define the oracle Bonferroni region
\begin{align}
 v^{\rm B}_{\psi,j}=\bar F_{\psi,j}^{-1}(1-\alpha/d),\qquad
 R^{\rm B}_\psi=\{x:\psi_j(x^{(j)})\geq v^{\rm B}_{\psi,j},\ j=1,\ldots,d\}.
 \label{vB}
\end{align}
In general, the union-bound argument implies
$v^F_\psi\geq\min_jv^{\rm B}_{\psi,j}$, but this inequality is insufficient to
order the two regions.  The inclusion
$R_\psi(v^F_\psi)\subseteq R^{\rm B}_\psi$ would follow from
$v^F_\psi\geq\max_jv^{\rm B}_{\psi,j}$, whereas the reverse inclusion would follow from
$v^F_\psi\leq\min_jv^{\rm B}_{\psi,j}$.  However, when the standardised lifetime margins agree, all marginal Bonferroni cutoffs coincide and the
first inclusion holds, as stated below.

\begin{proposition}\label{prop:bonf}
Suppose $\bar F_{\psi,j}=\bar F_{\psi,1}$ for every $j$ and this common survival
function is continuous.  Then the marginal Bonferroni cutoffs are equal to a common
$v^{\rm B}_\psi$, and
\begin{align}
 R_\psi(v^F_\psi)\subseteq R^{\rm B}_\psi=R_\psi(v^{\rm B}_\psi).
 \label{bonfcontainment}
\end{align}
Furthermore, $\bar F_\psi(v^{\rm B}_\psi)=1-\alpha$ if and only if the events
$\{\psi_j(X^{(j)})<v^{\rm B}_\psi\}$, $j=1,\ldots,d$, are pairwise disjoint up to null
sets.  If they are not, then $v^F_\psi>v^{\rm B}_\psi$.
\end{proposition}
Let $A_j=\{\psi_j(X^{(j)})<v^{\rm B}_\psi\}$.  The Bonferroni bound replaces the
joint miscoverage probability $\P(\bigcup_j A_j)$ by the upper bound
$\sum_j\P(A_j)=\alpha$; this inequality is strict exactly when two or more of the
events $A_j$ overlap with positive probability.  In the power-diagonal model of
Example~\ref{ex:exponential}, $v^{\rm B}_\psi=\kappa(\alpha/d)$ and
$v^F_\psi=\kappa(\alpha)/\theta_X$.  Using
$\kappa(a)=a+a^2/2+O(a^3)$ gives
\begin{align}
 \frac{v^F_\psi}{v^{\rm B}_\psi}
 =\frac{\kappa(\alpha)}{\theta_X\kappa(\alpha/d)}
 =\frac{d}{\theta_X}
   \{1+\tfrac{\alpha}{2}(1-1/d)+O(\alpha^2)\},\qquad \alpha\downarrow0,
 \label{gain}
\end{align}
so for fixed $d$ the ratio tends to $d/\theta_X$.  This limit is $1$ under independent
lifetimes and $d$ under comonotone lifetimes, so the gain from joint calibration grows
with the dependence between the lives.


\section{Selection and estimation of the path}\label{sec:design}
The prescribed coverage level can be attained by more than one orthant.  We first
specify an oracle criterion for selecting an orthant path and then study the effect of
estimating that path on the training fold.

\subsection{A simple selection rule for orthant corners}\label{sec:equivariance}
Let $\bar F_X(a)=\P(X\geq a)$ be the joint and $\bar F_j(t)=\P(X^{(j)}\geq t)$ the
marginal survival functions. Define
$\mathcal A_\alpha=\{a\in\R^d:\bar F_X(a)=1-\alpha\}$.  Every $a\in\mathcal A_\alpha$
satisfies
\begin{align}
 \bar F_j(a_j)\geq1-\alpha,\qquad j=1,\ldots,d .
 \label{feasible}
\end{align}
Assume that the marginal distribution functions are continuous and strictly increasing,
with support $\R$ for convenience, and let $\mathcal K_X$ be the survival copula of
$X$.  In the marginal survival coordinates $u_j=\bar F_j(a_j)$, exact coverage becomes
$\mathcal K_X(u)=1-\alpha$.  Selecting $u$ as a function of the copula and setting
$a_j=\bar F_j^{-1}(u_j)$ gives an equivariant rule, since a continuous increasing bijection of
each lifetime scale transforms the selected corner in the same way.  Equivariance alone
does not specify $u$.  The criterion below maximises the smallest marginal cumulative
hazard at the corner, $-\log u_j$.
\begin{lemma}[maximin corner]\label{lem:maximin}
Let $\Phi(u)=\mathcal K_X(u,\ldots,u)$ and let $u^*\in(0,1)$ satisfy $\Phi(u^*)=1-\alpha$
and $\Phi(u)<1-\alpha$ for $u<u^*$.  Then
\begin{align*}
 \max\bigl\{\min_j(-\log u_j):\ u\in(0,1)^d,\ \mathcal K_X(u)=1-\alpha\bigr\}=-\log u^*,
\end{align*}
and the diagonal vector $(u^*,\ldots,u^*)$ is a maximiser.
\end{lemma}

Notice that no symmetry of the copula is required.  The choice
\begin{align}
 \psi_j(t)=-\log\bar F_j(t),\qquad j=1,\ldots,d,
 \label{equalpath}
\end{align}
makes each $\psi_j(X^{(j)})$ standard exponential.  Along this equal-marginal path, the
cutoff $v^F_\psi=-\log u^*$ therefore defines a maximin corner, at which every life
has the same marginal survival probability $u^*$.  This is an unconditional
criterion; a conditional version replaces each marginal survival function by its
conditional counterpart and requires a covariate-dependent path.  We use the same path
for all contracts, and the covariates enter through the calibrated cutoff.

For example, if
$\bar F_j(t)=\exp\{-(t/\sigma_j)^{k_j}\}$ for $t\geq0$, then
$\psi_j(t)=(t/\sigma_j)^{k_j}$ and, for $v>0$,
\begin{align*}
 R_\psi(v)=\prod_{j=1}^d[\sigma_jv^{1/k_j},\infty).
\end{align*}
Thus one standardised cutoff corresponds to coordinate-specific bounds in the original time
scales.

\begin{figure}[!t]
\centering
\includegraphics[width=0.58\textwidth]{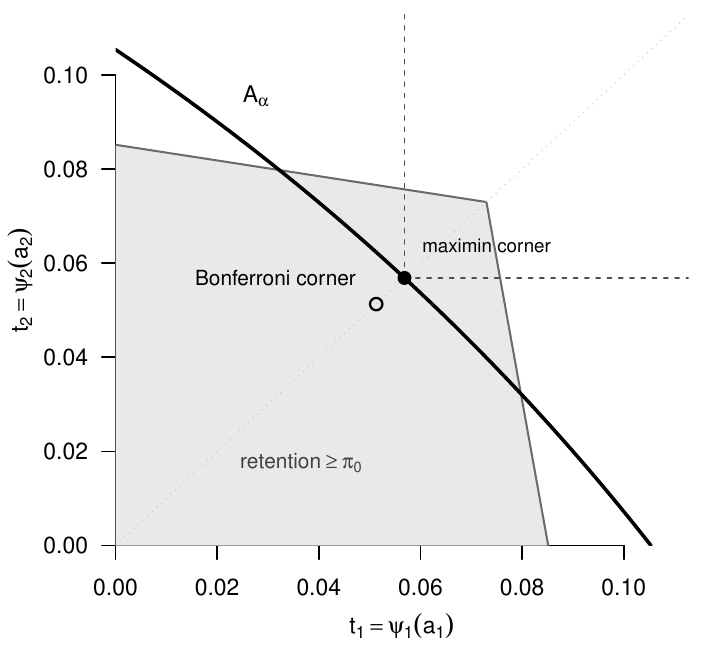}
\caption{Oracle corners on the equal-marginal scale, $t_j=\psi_j(a_j)=-\log\bar F_j(a_j)$,
with $d=2$ and $\alpha=0.1$.
The solid curve consists of the corners with coverage exactly $0.9$ for a Frank survival
copula of the lifetimes with Kendall's $\tau=0.4$, calculated numerically.  In the
shaded set, the retention probability is at least $\pi_0=0.6$
under the common-shock model of Example~\ref{ex:MO}, with $\lambda_0=5$ for the
valuation date and $\lambda_1=1$ for the coordinate-specific component.  The filled and
open points are the maximin and Bonferroni corners, respectively; the dashed lines
bound the maximin orthant.}
\label{fig:frontier}
\end{figure}

The maximin criterion does not account for censoring, and its selected corner may have
retention probability below a prescribed minimum $\pi_0$.  Figure~\ref{fig:frontier} illustrates
a case in which the criterion and the retention requirement are compatible.  Different
priorities among the lives could instead be represented by a weighted criterion on the
standardised coordinates.

\subsection{Stability under path estimation}\label{sec:pathstability}

The coverage in Theorem~\ref{thm:validity} holds for every
training-measurable path, so estimating the path does not affect it.  Consistency of the
path is needed only to compare the resulting region with that of the oracle path.  Fix a finite
horizon $c_0$, a positive margin $\delta>0$ and an oracle path $\psi$ with coordinates in
$(-\infty,\infty]$.  For $\varepsilon\geq0$, let $\mathcal N(\varepsilon)$ consist of all
paths $\varphi$ satisfying
\begin{align}
 \max_{j\leq d}\sup_{x:\psi_j(x)\leq c_0+\delta}
 |\varphi_j(x)-\psi_j(x)|\leq\varepsilon.
 \label{Nbhd}
\end{align}
with $|\pm\infty-a|=\infty$ for real $a$.  The margin $\delta$ above $c_0$ avoids
requiring accuracy over the whole support, and it may be taken small. This is important when
the observation period is bounded, since the oracle path can then be estimated only up
to the level of the observation period.

For any path $\varphi$, define
\begin{align*}
 q_\varphi(w)=\bar F_\varphi^{-1}(1-\alpha\mid w)\wedge c_0,
\end{align*}
conditioning additionally on $\Ztr$ when the path is fitted.  With the same
horizon $c_0$, the quantity $q(w)$ in Section~\ref{sec:dr} is simply
$q_{\hat\psi}(w)$.

\begin{assumption}[path estimation]\label{as:path}
The value $c_0+\delta$ belongs to the range of every $\psi_j$, and there is a
deterministic sequence $\varepsilon_N\downarrow0$ such that
\begin{align}
 \P(\hat\psi\in\mathcal N(\varepsilon_N))\longrightarrow1.
 \label{plugin}
\end{align}
\end{assumption}

In the following result, fitted-path probabilities condition on $\Ztr$, and all
statements involving $w$ are restricted to a common set of full $P_W$ measure.

\begin{theorem}[path stability]\label{thm:pathstability}
Suppose Assumption~\ref{as:path} holds, $q_\psi(w)$ is finite, and
$\sup_wq_\psi(w)<c_0$.  On $\{\hat\psi\in\mathcal N(\varepsilon_N)\}$, for all
sufficiently large $N$,
\begin{align}
 R_\psi(v+\varepsilon_N)&\subseteq R_{\hat\psi}(v)\subseteq R_\psi(v-\varepsilon_N),
 \qquad v\leq c_0,\label{pathsandwich}\\
 \bar G_\psi(c_0+\varepsilon_N)&\leq\bar G_{\hat\psi}(c_0)
 \leq\bar G_\psi(c_0-\varepsilon_N).\label{pathretention}
\end{align}
The conditional quantiles and their regions satisfy
\begin{align}
 \sup_w|q_{\hat\psi}(w)-q_\psi(w)|&\leq\varepsilon_N,\label{pathquantile}\\
 R_\psi(q_\psi(w)+2\varepsilon_N)&\subseteq
 R_{\hat\psi}(q_{\hat\psi}(w))\subseteq R_\psi(q_\psi(w)-2\varepsilon_N).
 \label{pathregion}
\end{align}
If Assumption~\ref{as:D} holds for the oracle path, then also
\begin{align}
 \sup_w\left|\P(X\in R_{\hat\psi}(q_{\hat\psi}(w))\mid W=w,\Ztr)
              -(1-\alpha)\right|\leq2b_2\varepsilon_N.
 \label{pathcoverage}
\end{align}
\end{theorem}

The set inclusions do not require smooth fitted paths.  For example, under the complete
independence condition in Assumption~\ref{as:coic}, suppose $X$ and $C$ are nonnegative with continuous margins,
and $\psi_j=-\log\bar F_j$ is strictly increasing on the relevant range.  If
$t_j=\psi_j^{-1}(c_0+\delta)$ is finite and $\P(Z^{(j)}\geq t_j)>0$, the marginal
Kaplan--Meier estimator has uniform $O_{\P}(N^{-1/2})$ error on $[0,t_j]$
\citep{BreslowCrowley1974}.
Since $\bar F_j\geq e^{-(c_0+\delta)}$ there, applying
$-\log$ preserves this rate, and for fixed $d$ the resulting path satisfies
\eqref{plugin} with the deterministic sequence $\varepsilon_N=N^{-1/2}\log N$.
This conclusion relies on complete independence.  Under the weaker conditional
independence condition in Assumption~\ref{as:cic}, $X$ and $C$ may be marginally
dependent through $W$, and unadjusted marginal Kaplan--Meier estimators need not be
consistent.

For Algorithm~\ref{alg:main}, suppose the assumptions of
Theorem~\ref{thm:dr}\,(ii) are satisfied for the fitted path.  Then
\eqref{qconsistency} and \eqref{pathquantile} imply
$\widehat L(W)-q_\psi(W)\to0$ in probability.  Moreover,
\eqref{pathsandwich} bounds the fitted region by oracle-path regions whose cutoffs
converge to $q_\psi(W)$.


\section{Simulation study}\label{sec:sim}
This section examines the finite-sample behaviour of the proposed procedure.

\subsection{Design}\label{sec:simdesign}

The experiments evaluate finite-sample coverage under independent censoring (setting S1),
retention depending on covariates (setting S2) and a short observation period (setting S3).  Separate diagnostic
experiments examine departures from conditional independence (setting S4) and the accuracy of the
estimated path (setting S5).  Table~\ref{tab:simdesign} summarises the settings, sample sizes and
replication counts.

\begin{table}[h]
\centering\small
\begin{tabular}{@{}lp{0.54\textwidth}rr@{}}
\hline
Setting & Description & Sample size & Replications \\
\hline
S1 & \raggedright Independent common censoring & 2000 & 500 \\
S2 & \raggedright Covariate-dependent retention and mortality hazards & 500/2000/8000 & 500/500/200 \\
S3 & \raggedright Empirical ages and censoring times; fitted lifetime model & 12000 & 300 \\
S4a, S4b & \raggedright Departures from conditional independence in S3; hazard multipliers $0.8/1.2$ for censoring times below five years & 12000 & 300 \\
S5 & \raggedright Path estimation under standardised exponential censoring & \begin{tabular}[t]{@{}r@{}}250/1000/\\4000/16000\end{tabular} & 200 \\
\hline
\end{tabular}
\caption{Simulation settings.  Sample size is the total $M=N+n$, split equally between
training and calibration, except in S5, where it is the training size $N$.
Replication counts apply separately to each sample size or variant; slash-separated
counts correspond to the listed sample sizes.}
\label{tab:simdesign}
\end{table}

In the setting S1, paired entry ages are sampled with replacement from their empirical
distribution in the data of Section~\ref{sec:data}, restricted to male entry ages of at least 70.  Remaining lifetimes have
Gompertz survival functions
\begin{align*}
 S_j(t\mid x)=\exp\{e^{(x-m_j)/\sigma_j}(1-e^{t/\sigma_j})\},
\end{align*}
with male and female modes $(86.4,92.2)$ and dispersions $(9.8,8.1)$ in years, motivated
by the univariate estimates of \citet{FreesCarriereValdez1996}.
Here $x$ is the entry age of the relevant life, with $j=1$ denoting the male
coordinate and $j=2$ the female coordinate.  A Frank survival copula
with Kendall's $\tau=0.34$ joins the conditional margins.  The common censoring time
$A\sim U(3,15)$ is independent of lifetimes and ages.  Its training-fold median defines
the horizon in years, $a_0$, approximately nine years, and its value on the path
scale, $c_0=\min_j\hat\psi_j(a_0)$, gives the retention event $\{A\geq a_0\}$.

In the S2 setting, $W=(W_1,W_2,W_3)$, where $(W_1,W_2)$ has the paired entry-age distribution
of S1 and $W_3\sim U(0,1)$ is independent.  The baseline Gompertz hazards and Frank
copula are as in S1, with both mortality hazards multiplied by
$\exp\{-2(W_3-\tfrac12)\}$.  Retention probabilities satisfy
\begin{align*}
 \P(A\geq9\mid W=w)=\mathrm{logit}^{-1}\{\mathrm{logit}(0.6)+4(w_3-\tfrac12)\}.
\end{align*}
Conditional on $W=w$, $A$ is uniform on $(9,15)$ with probability
$\P(A\geq9\mid W=w)$ and uniform on $(3,9)$ otherwise.  This probability ranges from
0.17 to 0.92 and increases with $w_3$, whereas the mortality-hazard multiplier decreases
with $w_3$.  Retained observations therefore have lower conditional mortality hazards
on average.  The horizon is $a_0=9$ years, and
$c_0=\min_j\hat\psi_j(a_0)$.

In the S3 setting, we sample entry ages and censoring times jointly from their empirical distribution and
use $a_0=5$ years.  Conditional lifetimes follow a Gompertz--Frank model fitted in advance
by censored maximum likelihood to the cleaned dataset of 12,357 observations.
In the parametrisation used in setting S1, its fitted male and female parameters are
$(m_1,\sigma_1)=(86.25,10.65)$ and $(m_2,\sigma_2)=(92.04,8.16)$, and its Kendall's
$\tau$ is $0.333$.

The settings S4a and S4b violate
conditional independence by multiplying the hazards of observations with $A<5$ by 0.8
and 1.2, respectively.  Setting S5 examines how path-estimation error affects the oracle
cutoff and the corners of the prediction region.  It keeps the entry-age and lifetime
distributions of S1 and uses the standardised exponential censoring model of
Example~\ref{ex:MO}, with $\lambda_0=\lambda_1=1$ and minimum retention probability $0.7$, to permit
direct comparison of the estimated and true paths.

In the settings S1 and S2, each coordinate $\psi_j=-\log\bar F_j$ of the equal-marginal path is estimated by Kaplan--Meier, linearly interpolated
between event times and simply extrapolated with its terminal secant slope.  Let
$\hat q_{\mathrm{AFT}}(\alpha\mid w)$ denote the conditional $\alpha$-quantile
of the reduced lifetime estimated by a Weibull accelerated-failure-time
(AFT) model and capped at $c_0$.  The corresponding
nonconformity score is
$V(w,y;\Ztr)=\hat q_{\mathrm{AFT}}(\alpha\mid w)-y$.
Furthermore, the setting S1 uses unit calibration weights, while setting S2 uses inverse-retention weights obtained from a
logistic regression, and with the estimated retention probabilities bounded below by 0.02.
To examine the double robustness result of Theorem~\ref{thm:dr}, setting S2 includes
two variants.  The first fits a single outcome quantile for all observations,
without using $W$, and applies either unit or fitted inverse-retention weights.  The
second uses the true conditional outcome quantile and unit weights.  Setting S3 uses the identity path together with the
Weibull AFT score and fitted logistic inverse-retention weights.

The reported benchmarks are the Bonferroni bounds with coordinate-specific
horizons $c_{0,j}=\hat\psi_j(a_0)$, so that every coordinate uses the retention event
$\{A\geq a_0\}$,
the naive calibration on $Z$, which treats censored lifetimes as deaths, and the oracle
region $R_{\hat\psi}(q_{\hat\psi}(w))$, based on the true conditional quantile along the
fitted path.  In setting S1 we also include a correctly-specified Gompertz--Frank model,
estimated by censored maximum likelihood on the training fold.

Coverage is computed from the true conditional distribution and averaged over 2,000
fixed test covariate vectors; in the settings S4a and S4b this distribution also conditions on the
censoring times.  This conditioning avoids variation from sampled test lifetimes but still
approximates integration over the covariate distribution.  All reported results use
$\alpha=0.1$.

\subsection{Results}\label{sec:simresults}
An overview of the simulation results is provided in Table~\ref{tab:sim}.

\begin{table}[h]
\centering\small
\begin{tabular}{lccccc}
\hline
Method & Coverage (sd) & Abs.\ dev. & Male & Female & At cap (\%) \\
\hline
\tableinput{tables/sim_main.tex}
\hline
\end{tabular}
\caption{Simulation results at $\alpha=0.1$.  Coverage is the conditional coverage
averaged over the fixed test covariates and replications; its standard deviation is
across replications.  Abs.\ dev.\ is the mean absolute conditional-coverage deviation
from $0.9$.  Male and Female are mean lower bounds in years.  At cap is the proportion of
the bounds at the horizon, using the male coordinate for Bonferroni; for the uncapped
Gompertz--Frank model it is the proportion at or above that level.  Settings S1 and S2 use the
estimated equal-marginal path, while setting S3 uses the identity path.}
\label{tab:sim}
\end{table}

In setting S1, where the censoring time is independent of lifetimes and covariates, the unweighted
joint region has coverage 0.904.  Its mean male and female lower bounds are 1.82 and 4.48
years, compared with
1.87 and 4.62 for the oracle along the same fitted path.  The bounds estimated under the
correctly specified Frank model are similar.  Bonferroni calibration is conservative,
whereas naive calibration gives coverage 0.922 and smaller coordinate bounds of 1.23
and 3.28 years.

Setting S2 aims to distinguish the retention and quantile models of Theorem~\ref{thm:dr}.  At $M=2000$, the
covariate-free score with unit weights has coverage 0.877.  Replacing the unit weights
by fitted logistic weights raises coverage to 0.902, while using the oracle conditional
quantile with unit weights gives coverage 0.907.  The corresponding values over all
three sample sizes are reported in Table~\ref{tab:simS2}.  About 10\% of the oracle
quantiles are capped, so the conditional limit in Theorem~\ref{thm:dr}(ii) does not
apply throughout the covariate space.  The procedure based on the fitted
accelerated-failure-time score also gives marginal coverage near 0.9.  The Weibull
AFT model need not consistently estimate the conditional quantiles generated by
the Gompertz model.  However, the logistic retention model is correctly specified in
setting S2, so Theorem~\ref{thm:dr}(i) gives asymptotic marginal validity without requiring a
consistent quantile model.

\begin{table}[h]
\centering\small
\begin{tabular}{rcccc}
\hline
& AFT score, & \multicolumn{2}{c}{Covariate-free score} & Oracle quantile, \\
$M$ & fitted weights & unit weights & fitted weights & unit weights \\
\hline
\tableinput{tables/sim_S2.tex}
\hline
\end{tabular}
\caption{Coverage in setting S2 when retention favours observations with lower mortality
hazards.  Parentheses contain standard deviations across replications.  The
covariate-free score with unit weights misspecifies both models.  The
fitted-weight and oracle-quantile columns satisfy the conditions of
Theorem~\ref{thm:dr}\,(i) and (ii), respectively.}
\label{tab:simS2}
\end{table}

Setting S3 uses an empirical censoring distribution with a short observation period.  With the
identity path, the fitted and oracle procedures have coverage 0.910 and 0.909 and mean
bounds of 3.85 and 3.87 years, respectively.  Bonferroni and naive calibration are more
conservative.  Table~\ref{tab:simS4} compares true population coverage with an estimate
that mimics the real-data evaluation.  This estimate uses only test observations with
$A\geq5$, for which membership in the prediction region is determined by the
observed censored outcomes.  Writing $I_i$ for this membership indicator and
$\hat\omega_i=1/\hat\pi(W_i)$, the estimate is $\sum_i\hat\omega_iI_i/\sum_i\hat\omega_i$,
with both sums over these retained test observations; it estimates population
coverage under conditional independence.  The discrepancies in the sensitivity analyses of settings S4a and S4b arise
because their mortality hazards depend on $A$ even after conditioning on the
covariates.

\begin{table}[h]
\centering\small
\begin{tabular}{lccc}
\hline
Setting & Hazard multiplier & Target coverage & Weighted estimate \\
\hline
\tableinput{tables/sim_S4.tex}
\hline
\end{tabular}
\caption{Sensitivity of the observable weighted coverage estimate in the data-calibrated setting
S3 and the variant settings S4a and S4b at $M=12000$.  Target coverage is computed from the
simulated lifetime distribution; the estimate uses retained test observations and fitted
inverse-retention weights.
Setting S3 correctly satisfies conditional independence, whereas settings S4a and S4b multiply the hazards when
$A<5$ as shown.  The mean standard error of the estimate rounds to 0.008 in each row.}
\label{tab:simS4}
\end{table}

Finally, in setting S5, the average maximum path error over the evaluation grid decreases from 0.042 to
0.005 as the training size increases from 250 to 16000, with a log--log slope of
$-0.49$.  Table~\ref{tab:simS5} also reports the corresponding quantile and corner
errors.  These results are in line with root-$N$ stochastic error rates.

\begin{table}[h]
\centering\small
\begin{tabular}{lcccc}
\hline
$N$ & Path error & Quantile error & Corner, mean & Corner, 95\% \\
\hline
\tableinput{tables/sim_S5.tex}
\hline
\end{tabular}
\caption{Path estimation in setting S5, with minimum retention probability $\pi_0=0.7$, margin
$\delta=0.05$ and $\alpha=0.1$.  Path error is the maximum absolute estimation error
across coordinates, evaluated for each coordinate on a 400-point grid over
$\{t\geq0:\psi_j(t)\leq c_0+\delta\}$.  For the fixed test covariates
$w_k$, quantile error is $\max_k|q_{\hat\psi}(w_k)-q_\psi(w_k)|$.  Corner displacement
is the absolute difference, in months, between the first (male) coordinates of the
corresponding fitted- and oracle-path corners, summarised by its mean and 95th
percentile over the test covariates.  Entries are averaged over replications.  In
every replication, $\bar G_{\hat\psi}(c_0)$ lay between the two bounds in
\eqref{pathretention}.  The three retention probabilities were approximated from the
same 200,000 draws from the censoring distribution.  In each replication,
$\varepsilon_N$ was the maximum absolute difference between the estimated and true
paths over the grid defined above.}
\label{tab:simS5}
\end{table}

\FloatBarrier
\section{Joint-life annuity data}\label{sec:data}
The application estimates lower prediction bounds for reported first- and second-death times
in a joint-life annuity portfolio with a short observation period.

\subsection{Data}\label{sec:datadesc}

The Canadian insurance data of \citet{FreesCarriereValdez1996}, found in
CASdatasets \citep{DutangCharpentier2026}, contain 14,889 male--female annuity contracts
observed from 29 December 1988 to 31 December 1993.  Entry ages and reported death
times are available.  For every contract, the file also records the length $A$ of its
observation period, from entry into observation to 31 December 1993, which equals the study
length of 5 years for 69\% of the 14,889 contracts and is shorter for individuals entering later. We take
$C=(A,A)$ as the observed administrative censoring vector.

We remove 2,529 duplicates and three observations with entry age below 18, which leaves
a sample of 12,357 observations.  The remaining rows are treated as distinct
observations, since the file has no identifiers with which to detect repeated lives;  repeated
lives could produce dependence.  Median male and female entry ages are
67.6 and 65.2 years.
When observations are grouped by male and female entry age, the proportions satisfying
$A\geq5$ rise from 49.6\% and 55.6\% below age 65 to 83.0\% and 82.3\% at ages
75--80, respectively
(Table~\ref{tab:fourstate}).  This age dependence motivates
estimating the retention probability $\pi(w)=\P(A\geq5\mid W=w)$ in
\eqref{weights} from the entry ages and using inverse-retention weights
$1/\hat\pi(w)$ in calibration.

\begin{table}[H]
\centering\small
\begin{tabular}{lrrrr}
\hline
Entry-age band & $n$ & At least one death & Both deaths & $A\geq5$ (\%) \\
\hline
\tableinput{tables/data_fourstate.tex}
\hline
\end{tabular}
\caption{Sample composition by male and female entry age.}
\label{tab:fourstate}
\end{table}

The analysis is carried out with reported death times rather than verified biological death times or
payment-cessation times.  As noted in \citet{FreesCarriereValdez1996}, a death during a
guarantee period, or the first death under a joint-and-last-survivor annuity, may leave
payments unchanged and therefore go unreported.  Both deaths are reported for 198
observations in the cleaned sample.  Their reported times differ by at most one year in
128 cases, including 56 cases in which they differ by at most one week.  These counts are
descriptive and one cannot know whether reporting affected the observed
gaps. 
\subsection{Design}\label{sec:datadesign}
The population consists of couples with both lives alive when the observation period begins,
and we treat the two entry ages as covariates.  The remaining lifetimes are measured from that
date onwards.  The conclusions are therefore conditional on survival of both lives to entry and
do not apply to couples that might have been excluded due to left truncation.  We repeat five-fold sample splitting
20 times.  For each split, every fold serves once as the test set, and the remaining four
folds are divided evenly into training and calibration sets.

We consider $\psi_1(x)=\psi_2(x)=x$, so the reduced outcome is
$\min(X^{(1)},X^{(2)})$.  The calibrated cutoff $L$ defines the region
$[L,\infty)^2$ and is therefore a lower prediction bound for the reported first-death
time.  With the identity path, the horizon is $c_0=a_0$, and we set $a_0=5$ years; 65\% of
the 12,357 observations satisfy the resulting retention event.  Conditional quantiles in the score are fitted by a Weibull
accelerated-failure-time regression on both entry ages.  Logistic regression on the same
ages estimates the retention probabilities, which are bounded below by 0.02.
Both models are fitted only on the training fold.  Although both entry ages already
enter these models as covariates, we also compute the conformal adjustment $\eta(w)$
separately within the male entry-age bands and, in a separate analysis, within the female
entry-age bands of Table~\ref{tab:fourstate}.  Since coverage holds only on average over
ages, it may vary across bands; hence calibrating within a band targets coverage in that specific band
(Section~\ref{sec:validity}).  The
validity result with estimated weights is asymptotic and requires the conditions of
Section~\ref{sec:dr} to hold; correspondingly, bandwise calibration requires the same conditions within each
band.

The Bonferroni benchmark
calibrates the two lifetimes separately.  Before calibration, each lifetime is put on a
marginal cumulative-hazard scale using its training-fold Kaplan--Meier estimate, with
linear interpolation between event times.  A coordinate-specific Weibull
accelerated-failure-time score is then calibrated at miscoverage level $\alpha/2$, using
the same retained sample $\{A\geq a_0\}$ and inverse-retention weights as in the primary
procedure.  The two lower bounds are then transformed back to years and intersected to form
the joint region.  Two further benchmarks are used for the first-death bound.  The
naive benchmark calibrates one bound for all ages directly on $Z$, without retention weights.  The
parametric benchmark fits Gompertz marginal distributions and a Frank survival copula
to the two remaining lifetimes, conditional on entry ages, using all non-test
observations\footnote{ This benchmark is not a direct reproduction of the model in
\citet{FreesCarriereValdez1996}.  That model specifies the joint distribution of ages
at death and then conditions on both lives having survived to entry, whereas the
benchmark applies the Frank copula directly to the conditional remaining lifetimes.}.  The fitted $\alpha$-quantile of $\min(X^{(1)},X^{(2)})$ is its lower
first-death bound.

The identity-path reduction above gives the minimum of the two lifetimes and is used
only for the first-death analysis.  For the secondary analysis, we instead use the
common-date reduction $s(x)=\max(x^{(1)},x^{(2)})$ from
Proposition~\ref{prop:common}.  Since $s(Z)=s(X)\wedge A$, the reported second-death
time $s(X)$ is an ordinary single lifetime right censored at the observed time $A$.
At nominal level 0.95, we apply the same type of Weibull accelerated-failure-time
score, horizon and estimated weights as in the primary procedure.  The
higher coverage level lowers the target quantile and reduces capping for the later
second-death time.  We compare calibration pooled over all ages with calibration within the two oldest male
entry-age bands and, separately, within the two oldest female entry-age bands.

For a retained test observation, $A\geq a_0$, and every evaluated coordinatewise
lower bound $L_j$ is at most $a_0$.  Hence $X^{(j)}\geq L_j$ if and only if
$Z^{(j)}\geq L_j$, so the data determine whether $X^{(j)}\geq L_j$ for every
coordinate.  Likewise, a second-death bound $L\leq a_0$ satisfies $s(X)\geq L$ if and
only if $s(Z)\geq L$.  Tables~\ref{tab:bands} and~\ref{tab:seconddeath} report the
resulting proportions of retained test observations in the region, which we call
empirical inclusion.  We also
compute the inverse-retention-weighted ratio of Section~\ref{sec:simresults}, which
estimates marginal coverage in the target population if Assumption~\ref{as:cic} holds
and the retention probabilities are estimated consistently. 

For each selected age pair, each five-fold repetition produces one bound from each test
fold.  We average these five bounds and report the median of the 20 repetition-specific
averages.

\subsection{Results}\label{sec:dataresults}
We report lower bounds for the first- and second-death times separately.  In each
analysis, empirical inclusion is reported together with the frequency at which the
fitted bound reaches the five-year cap, since capping can increase inclusion and
prevents evaluation of lower bounds above five years.

\subsubsection{First-death bounds}
At nominal level 0.9, the median unweighted empirical inclusion probability among retained test
observations is 0.914, and the corresponding inverse-retention-weighted estimate is
0.918.

Panel~(a) in Table~\ref{tab:bands} shows that the pooled overall inclusion probability combines rather different age-band
results.  In the youngest bands, the bound is fixed at five years for 97.0\% of
male-band observations and 61.7\% of female-band observations, so these inclusion
proportions primarily reflect the imposed cap.  In contrast, capping is negligible in the oldest
bands, where pooled inclusion probability falls to 0.874 and 0.860.  Bandwise calibration lowers
the corresponding mean bounds from 1.03 to 0.70 years and from 1.00 to 0.63 years,
raising inclusion probabilities to 0.910 and 0.915.  In the 75--80 bands it instead raises the mean
bounds and lowers inclusion probabilities.  Thus, we see that restriction of calibration to an age band changes
the results even after fitting the accelerated-failure-time quantile model.  The oldest bands
contain only 277 and 164 retained observations, so these empirical proportions are
less stable.

Panel~(b) in Table~\ref{tab:bands} gives the corresponding results for the benchmarks, also by age band.  Its implied inclusion probability falls to 0.783
and 0.741 in the oldest bands.  The Bonferroni inclusion probability remains above 0.9 through ages 75--80 but
falls below 0.9 in the oldest bands.  The Frank benchmark gives inclusion close to that of the pooled procedure in each age
band.  Variation across age bands therefore also appears
under the parametric model.

The Frank copula fitted to the full sample has Kendall's $\tau=0.33$.  It gives an expected 175.3
observations in which both deaths occur before the recorded censoring time, compared
with the actual 198 observed and with 81.8 under an independence model with the same fitted marginal lifetime
distributions.  This comparison supports allowing positive dependence in the benchmark
but does not assess predictive performance on new observations.

The primary bound equals five years for 28\% of retained observations.
Reducing the horizon from five to three years raises
retention probability from 0.65 to 0.88, but the proportion of bounds at the horizon rises to
71\%.  The larger retained sample therefore comes at the cost of less informative
bounds.  At entry ages 70 and 67, the median primary and Frank bounds are 3.45 and 3.51
years.  At ages 65 and 62, the primary bound is capped at five years and the Frank
bound exceeds the observation period, so neither method can be evaluated beyond five
years using these data.

\begin{table}[H]
\centering\footnotesize
\textit{(a) Proposed procedures}\par\smallskip
\begin{tabular}{@{}lrcccccc@{}}
\hline
Entry-age band & Retained ($A\geq5$) & \multicolumn{2}{c}{Empirical inclusion}
& \multicolumn{2}{c}{At five-year cap (\%)}
& \multicolumn{2}{c}{Mean lower bound (years)} \\
\cline{3-8}
& & Pooled & Bandwise & Pooled & Bandwise & Pooled & Bandwise \\
\hline
\tableinput{tables/data_bands.tex}
\hline
\end{tabular}
\medskip
\textit{(b) Benchmarks}\par\smallskip
\begin{tabular}{@{}lccc@{}}
\hline
Entry-age band & \multicolumn{3}{c}{Empirical inclusion} \\
\cline{2-4}
& Bonferroni & Naive & Frank \\
\hline
\tableinput{tables/data_benchmarks.tex}
\hline
\end{tabular}
\caption{First-death analysis at nominal level 0.9.  Retained gives the number of
observations with $A\geq5$; all other numerical entries are medians across 20 repeated
five-fold partitions.  In panel (a), empirical inclusion is the unweighted
proportion of retained test observations satisfying $Z^{(1)}\geq L$ and
$Z^{(2)}\geq L$.  The cap percentage and mean lower bound are computed among the same
observations.  Pooled uses retained calibration observations from all age bands.
Bandwise restricts conformal calibration and evaluation to the indicated band; the
fitted quantile and retention models are unchanged.  Panel (b) reports unweighted
empirical inclusion for three benchmarks calibrated without age-band stratification
and evaluated in the indicated groups.  All lower bounds are capped at five years by design.}
\label{tab:bands}
\end{table}

\subsubsection{Second-death bounds}
The analysis based on
$s(X)=\max(X^{(1)},X^{(2)})$ is given in Table~\ref{tab:seconddeath}.  Pooling the conformal calibration over all ages gives
median unweighted empirical inclusion probability of 0.983 and a corresponding
inverse-retention-weighted estimate of 0.985, both above the nominal level 0.95.  The
fitted bound equals five years for 93\% of retained observations, and its mean is 4.9
years.  As before, the data say nothing about bounds above five years.  We report bandwise calibration only for entry ages of at least
75.  In the younger bands, between 93.4\% and 100\% of the pooled bounds equal five
years, leaving little variation to compare within the observation period.

\begin{table}[H]
\centering\footnotesize
\begin{tabular}{@{}lrcccccc@{}}
\hline
Entry-age band & Retained ($A\geq5$) & \multicolumn{2}{c}{Empirical inclusion}
& \multicolumn{2}{c}{At five-year cap (\%)}
& \multicolumn{2}{c}{Mean lower bound (years)} \\
\cline{3-8}
& & Pooled & Bandwise & Pooled & Bandwise & Pooled & Bandwise \\
\hline
\tableinput{tables/data_seconddeath.tex}
\hline
\end{tabular}
\caption{Second-death analysis at nominal level 0.95.  Retained gives the number of
observations with $A\geq5$; all other numerical entries are medians across 20 repeated
five-fold partitions.  Empirical inclusion is the
unweighted proportion of retained test observations satisfying
$\max\{Z^{(1)},Z^{(2)}\}\geq L$.  The cap
percentage and mean lower bound are computed among the same observations.  Pooled uses
retained calibration observations from all age bands.  Bandwise restricts conformal
calibration and evaluation to the indicated band; the fitted quantile and retention
models are unchanged.  All lower bounds are capped at five years by design.}
\label{tab:seconddeath}
\end{table}

Within the same age bands, bandwise calibration produces lower bounds that are less
often capped.  In the male 75--80 band, the proportion at five years decreases from
67.8\% to 29.0\%; in the female 75--80 band it decreases from 35.7\% to 6.3\%.  In the oldest
bands, the bandwise bounds are essentially never capped and their means are about one
year below those from pooled calibration.  Empirical inclusion probabilities increase
in all four bands; in the oldest male and female bands they change from 0.935 to 0.960 and from
0.951 to 0.963, respectively.  The
unweighted and inverse-retention-weighted estimates of the inclusion probability are nearly identical
within each band, so reweighting the retained test observations has little effect on
the within-band estimates.

For entry ages 80 and 77, the median fitted bound is 2.88 years under male-age
calibration and 3.36 years under female-age calibration; at ages 82 and 79, the
corresponding values are 2.17 and 2.64 years.  The difference reflects the distinct
sets of observations used for male- and female-band calibration.  Each procedure
calibrates within one age band at a time and does not estimate a second-death quantile
conditional on the two exact entry ages.

\section*{Acknowledgements}

The author was supported by the Carlsberg Foundation, grant CF23-1096.  The data are
distributed in the R package CASdatasets. Open AI ChatGPT-6 Astra and Claude Fable 5.1 were used to help with writing, coding and finding missing arguments in earlier versions of the proofs. The author has verified and curated the output from these models.

\newpage
\appendix

\section{Proofs}\label{app:proofs}

\subsection*{Proofs for Section~\ref{sec:setup}}
\begin{proof}[Proof of Theorem~\ref{thm:shape}]
To prove (ii) from (i), first note that if $x\leq y$ componentwise then $x=x\wedge y$, so \eqref{compat}
gives $s(x)=s(x)\wedge s(y)\leq s(y)$.  Thus $s$ is nondecreasing.

For $j\leq d$ and
$t,m\in\R$, let $e_j(t,m)\in\R^d$ have $j$th coordinate $t$ and all other coordinates
$m$.  Fix $x\in\R^d$ and $m\geq\max_jx^{(j)}$.  The $l$th coordinate of
$\bigwedge_{j=1}^de_j(x^{(j)},m)$ is $\min\{x^{(l)},m\}=x^{(l)}$, the value $x^{(l)}$
coming from the term $j=l$ and the value $m$ from the other terms; hence
$x=\bigwedge_je_j(x^{(j)},m)$.  Identity \eqref{compat} extends to finitely many
arguments by induction, so
\begin{align*}
 s(x)=\min_{1\leq j\leq d}s\bigl(e_j(x^{(j)},m)\bigr)\qquad\text{for every }m\geq\max_jx^{(j)}.
\end{align*}
For fixed $t$ the map $m\mapsto s(e_j(t,m))$ is nondecreasing, so
\begin{align*}
 \psi_j(t)=\lim_{m\to\infty}s(e_j(t,m))
\end{align*}
exists in $\bar\R$.  Each $\psi_j$ is nondecreasing because $t\mapsto s(e_j(t,m))$ is nondecreasing.
The left-hand side of the preceding identity for $s(x)$ does
not depend on $m$; letting $m\to\infty$ on the right, where the minimum of finitely many
arguments is continuous on $\bar\R^d$, gives \eqref{minform}.

To prove (i) from (ii), note that a nondecreasing $\psi:\R\to\bar\R$ satisfies
$\psi(a\wedge b)=\psi(a)\wedge\psi(b)$.  Hence
\begin{align*}
 s(x\wedge c)=\min_j\bigl\{\psi_j(x^{(j)})\wedge\psi_j(c^{(j)})\bigr\}
 =\Bigl(\min_j\psi_j(x^{(j)})\Bigr)\wedge\Bigl(\min_j\psi_j(c^{(j)})\Bigr)=s(x)\wedge s(c).
\end{align*}

To prove (iii) from (ii), write $\{s\geq v\}=\prod_j\{t\in\R:\psi_j(t)\geq v\}$.  Each
factor is an up-set of $\R$ because $\psi_j$ is nondecreasing.

To prove (i) from (iii), first note that a product of up-sets is an up-set of $\R^d$, so $x\leq y$ and
$s(x)\geq v$ imply $s(y)\geq v$ for every real $v$; hence $s$ is nondecreasing, and
$s(x\wedge c)\leq s(x)\wedge s(c)$.

For the reverse inequality let $v\in\R$ satisfy
$v\leq s(x)\wedge s(c)$.  Then $x$ and $c$ belong to $\prod_jI_j(v)$, so $x^{(j)}$ and
$c^{(j)}$ belong to $I_j(v)$ for every $j$; an up-set of $\R$ contains the smaller of any
two of its points, so $x^{(j)}\wedge c^{(j)}\in I_j(v)$ and $x\wedge c\in\{s\geq v\}$.
Thus $s(x\wedge c)\geq v$ for every real $v\leq s(x)\wedge s(c)$, which gives
$s(x\wedge c)\geq s(x)\wedge s(c)$, also when the right-hand side is $\pm\infty$.

For (iv), monotonicity follows from (i) as shown above.  If $s(c)\geq c_0$, then
$s(x\wedge c)\wedge c_0=s(x)\wedge s(c)\wedge c_0=s(x)\wedge c_0$.

To prove (i) from (iv), fix $x,c\in\R^d$.  If $s(c)$ is finite, take $c_0=s(c)$ in
\eqref{retention} to obtain $s(x\wedge c)\wedge s(c)=s(x)\wedge s(c)$, and
$s(x\wedge c)\leq s(c)$ by monotonicity, so the left-hand side equals $s(x\wedge c)$.

If $s(c)=+\infty$, the identity holds for every real $c_0$, and letting $c_0\to\infty$
gives $s(x\wedge c)=s(x)=s(x)\wedge s(c)$.  
If $s(c)=-\infty$, monotonicity gives
$s(x\wedge c)\leq s(c)=-\infty$.  In all cases \eqref{compat} holds.
\end{proof}

\begin{proof}[Proof of Proposition~\ref{prop:common}]
On $\{C=A\mathbf 1\}$, identity \eqref{diagcompat} with $t=A$ gives
\begin{align*}
 s(Z)=s(X\wedge A\mathbf 1)=s(X)\wedge s(A\mathbf 1)=s(X)\wedge s(C).
\end{align*}
Taking the minimum with $c_0$ gives the retention identity on $\{s(C)\geq c_0\}$.
\end{proof}

\subsection*{Proofs for Section~\ref{sec:method}}

\begin{proof}[Proof of Theorem~\ref{thm:reduction}]
Write $E=\{\rC\geq c_0\}$.  On $E$, Theorem~\ref{thm:shape} gives
$\rZ\wedge c_0=\rX\wedge\rC\wedge c_0=\rX\wedge c_0=Y$.  For \eqref{shift}, note that
$Y$ is a function of $X$ and $\ind_E$ is a function of $C$, so Assumption~\ref{as:cic}
gives $Y\indep\ind_E\mid W$.  Hence, for Borel sets $B\subseteq\overline\R$ and
$D\subseteq\R^p$,
\begin{align*}
 \P(Y\in B,\ W\in D,\ E)
 &=\E\bigl[\ind\{W\in D\}\,\P(Y\in B\mid W)\,\P(E\mid W)\bigr]\\
 &=\int_D\P(Y\in B\mid W=w)\,\P(E\mid W=w)\,P_W(dw).
\end{align*}
Dividing by $\P(E)=\bar G_\psi(c_0)>0$ and noting that
$\P(E\mid W=w)P_W(dw)/\P(E)$ is $P_{W\mid E}(dw)$ gives
\begin{align*}
 \P(Y\in B,W\in D\mid E)=\int_D\P(Y\in B\mid W=w)P_{W\mid E}(dw),
\end{align*}
which is \eqref{shift}.

The same computation gives the covariate density as
\begin{align*}
 \frac{dP_{W\mid E}}{dP_W}(w)
 =\frac{\P(E\mid W=w)}{\P(E)}
 =\frac{\bar G_\psi(c_0\mid w)}{\bar G_\psi(c_0)}.
\end{align*}
When
$\bar G_\psi(c_0\mid W)>0$ almost surely this density is positive, so $P_W\ll P_{W\mid E}$
and \eqref{LR} is its reciprocal.  Under Assumption~\ref{as:coic}, $E$ is independent of
$W$, so $\bar G_\psi(c_0\mid w)=\bar G_\psi(c_0)$ for $P_W$-almost every $w$.
\end{proof}

\begin{proof}[Proof of Theorem~\ref{thm:validity}]
Condition on $\Ztr$, which fixes $\hat\psi$, $c_0$, $V$ and $\omega$, and on the
retention indicators $\Delta_i=\ind\{\hat s(C_i)\geq c_0\}$, $i\in\Ica$, hence on $m=|\Icap|$
and on which observations are retained.  The calibration triples $(W_i,C_i,X_i)$ are
i.i.d.\ and independent of $\Ztr$.  If $\P(\hat s(C)\geq c_0\mid\Ztr)=0$, which can occur
only in part (i), then $m=0$ almost surely and the output is $\R^d$, which clearly attains the required coverage.

Otherwise, given the indicators,
the retained pairs $(W_i,Y_i)$, $i\in\Icap$, are independent, each distributed as $(W,Y)$
conditionally on $\{\hat s(C)\geq c_0\}$, which by \eqref{shift} is the distribution
$P'\times P_{Y\mid W}$ with $P'=P_{W\mid\hat s(C)\geq c_0}$.  The new observation $(W,Y)$ is
independent of the calibration fold and has distribution $P_W\times P_{Y\mid W}$.

We use the covariate-shift result of \citet[Corollary~1 and Section~2.2]{TibshiraniEtAl2019}
for split conformal prediction in the following form.  If calibration pairs are i.i.d.\ from $P'\times P_{Y\mid W}$, the
test pair is independent of them with distribution $Q\times P_{Y\mid W}$, $Q\ll P'$ with
$dQ/dP'=\omega$ up to a positive constant, and $\eta$ is the quantile of step~4 computed
with weights $\omega$, then $\P(V(W,Y;\Ztr)\leq\eta(W))\geq1-\alpha$.  The constant is
immaterial because $Q_w$ in step~3 is normalised.

Under Assumption~\ref{as:coic}, the indicators $\Delta_i$ are functions of the $C_i$ and
independent of $(W_i,X_i)$, so $P'=P_W$ and the theorem applies with $\omega\equiv1$.
Under Assumption~\ref{as:cic} with positivity, Theorem~\ref{thm:reduction} gives
$dP_W/dP'=\P(\hat s(C)\geq c_0\mid\Ztr)/\pi$, which is proportional to $1/\pi$, and the
theorem applies with
$\omega(w)=1/\pi(w)$.  In both cases
\begin{align*}
 \P(V(W,Y;\Ztr)\leq\eta(W)\mid\Ztr,(\Delta_i)_{i\in\Ica})\geq1-\alpha.
\end{align*}
If $m=0$ then $Q_w=\delta_{+\infty}$, $\eta(w)=+\infty$, and the event holds trivially.

Since $V(w,\cdot;\Ztr)$ is nonincreasing, the set $\{y:V(w,y;\Ztr)\leq\eta(w)\}$ is an
up-set of $\R$.  On the event $V(W,Y;\Ztr)\leq\eta(W)$, this set contains $Y$, and
hence its infimum is at most $Y$.  Since $Y\leq c_0$, the definition of
$\widehat L(W)$ gives $\widehat L(W)\leq Y$.  If $Y=-\infty$ is
possible on the path, then every real $y$ belongs to the set and the infimum is
$-\infty$, so the implication is preserved.  Thus, since $Y\leq\hat s(X)$,
\begin{align*}
 \{V(W,Y;\Ztr)\leq\eta(W)\}
 &\subseteq\{\widehat L(W)\leq Y\}\\
 &\subseteq\{\widehat L(W)\leq\hat s(X)\}
 =\{X\in\widehat R(W)\}.
\end{align*}
The last equality follows from \eqref{orthant}.  The conditional coverage is therefore
at least $1-\alpha$.  Averaging over the
indicators and over $\Ztr$ proves both parts.
\end{proof}

\begin{proof}[Proof of Theorem~\ref{thm:dr}]
Throughout, we condition on $\Ztr$.  All constants below depend only on $\alpha$, $\rho$,
$M$, $b_1$, $b_2$ and $r$, and not on the training fold, which permits averaging over
$\Ztr$ at the end.  Write $P$ for the distribution of $W$ in the target population and $P'$ for
its conditional distribution among retained observations, let $E=\{\hat s(C)\geq c_0\}$ be the retention event,
and let $\bar\pi=\P(E\mid\Ztr)=\E_P[\pi(W)\mid\Ztr]$ be its associated probability.  By
Theorem~\ref{thm:reduction}, $dP'/dP=\pi/\bar\pi$.

\emph{Step 1: normalisation and moments.}  Put
\begin{align*}
 A=\E_{P'}[\hat \pi(W)^{-1}\mid\Ztr],\qquad
 B=\E_{P'}[\pi(W)^{-1}\mid\Ztr]=1/\bar\pi,
\end{align*}
where the last equality follows from $dP'/dP=\pi/\bar\pi$.  Define
\begin{align*}
 a(w)=\frac{1}{A\,\hat \pi(w)},\qquad b(w)=\frac{\bar\pi}{\pi(w)},
\end{align*}
which both integrate to one under $P'$.  Since $A$ is a constant given $\Ztr$ and $Q_w$
is invariant to multiplying all weights by a constant, Algorithm~\ref{alg:main} with
$\omega(w)=1/\hat\pi(w)$ coincides with Algorithm~\ref{alg:main} with $\omega=a$.  By
Jensen's inequality and Assumption~\ref{as:O},
\begin{align*}
 \bar\pi^{-1-\rho}
 =(\E_P[\pi(W)\mid\Ztr])^{-1-\rho}
 \leq\E_P[\pi(W)^{-1-\rho}\mid\Ztr]\leq M.
\end{align*}
Hence $\bar\pi\geq\pi_0:=M^{-1/(1+\rho)}$.  Since $\hat \pi\leq1$, then $A\geq1$; and since $\pi\leq1$,
\begin{align}
 \E_{P'}[a(W)^{1+\rho}\mid\Ztr]&\leq\E_{P'}[\hat \pi(W)^{-1-\rho}\mid\Ztr]
 =\frac{1}{\bar\pi}\E_P[\hat \pi(W)^{-1-\rho}\pi(W)\mid\Ztr]\leq\frac{M}{\pi_0},\nonumber\\
 \E_P[a(W)^{1+\rho}\mid\Ztr]&\leq\E_P[\hat \pi(W)^{-1-\rho}\mid\Ztr]\leq M.
 \label{weightmoments}
\end{align}
Given $\Ztr$, each calibration observation is retained independently with probability
$\bar\pi$, so $m=|\Icap|$ is binomial with parameters $(n,\bar\pi)$; by Chebyshev's inequality
and $\bar\pi\geq\pi_0$,
\begin{align*}
 \P(m\leq n\pi_0/2\mid\Ztr)
 &\leq\P(|m-n\bar\pi|\geq n\pi_0/2\mid\Ztr)\leq\frac{4\bar\pi(1-\bar\pi)}{n\pi_0^2}\leq\frac{1}{n\pi_0^2}.
\end{align*}
Thus $m\to\infty$ in probability uniformly in the training fold.

\emph{Step 2: part (i).}  Let $\widetilde P$ be the probability measure with
$d\widetilde P/dP'=a$.  As in the proof of Theorem~\ref{thm:validity}, given $\Ztr$ and
the retention indicators the retained pairs are i.i.d.\ from $P'\times P_{Y\mid W}$, and
$a$ is the likelihood ratio of $\widetilde P$ against $P'$; the weighted split-conformal
theorem therefore gives $\P(V(W,Y;\Ztr)\leq\eta(W))\geq1-\alpha$ for a test pair with
distribution $\widetilde P\times P_{Y\mid W}$, hence $\P(Y\geq\widehat L(W))\geq1-\alpha$ under
that same distribution by the monotonicity argument of Theorem~\ref{thm:validity}.

We now transfer this guarantee to the target distribution.  The actual test
pair has distribution $P\times P_{Y\mid W}$, where $dP/dP'=b$.  The event $\{Y\geq\widehat L(W)\}$
is determined by the calibration fold and the test pair, and the two joint distributions of these
objects share the same distribution of the calibration fold and the kernel $P_{Y\mid W}$. Hence, the two
coverage probabilities differ by at most the total variation distance between
$\widetilde P$ and $P$, which is $\tfrac12\E_{P'}[|a(W)-b(W)|\mid\Ztr]$.

To bound this distance, write
$a-b=(\hat \pi^{-1}-\pi^{-1})/A+\pi^{-1}(A^{-1}-B^{-1})$.
Using $\E_{P'}[\pi^{-1}\mid\Ztr]=B$ gives
\begin{align*}
 \E_{P'}[|a-b|\mid\Ztr]
 \leq\frac{\E_{P'}[|\hat \pi^{-1}-\pi^{-1}|\mid\Ztr]}{A}+\frac{B\,|B-A|}{AB}
 \leq2\,\E_{P'}[|\hat \pi^{-1}-\pi^{-1}|\mid\Ztr],
\end{align*}
since $|B-A|=|\E_{P'}[\pi^{-1}-\hat \pi^{-1}\mid\Ztr]|\leq\E_{P'}[|\hat \pi^{-1}-\pi^{-1}|\mid\Ztr]$
and $A\geq1$.  Changing measure and using $\pi\leq1$, we obtain that 
$\E_{P'}[|\hat \pi^{-1}-\pi^{-1}|\mid\Ztr]\leq\pi_0^{-1}\E_P[|\hat \pi^{-1}-\pi^{-1}|\mid\Ztr]$.
Therefore
\begin{align*}
 \P(Y\geq\widehat L(W)\mid\Ztr)\geq1-\alpha-\pi_0^{-1}\E_P[|\hat \pi(W)^{-1}-\pi(W)^{-1}|\mid\Ztr].
\end{align*}
Since $\hat s(X)\geq Y$ and $\{\hat s(X)\geq\widehat L(W)\}=\{X\in\widehat R(W)\}$,
averaging over $\Ztr$ and applying Assumption~\ref{as:C} proves (i).  Note that nothing in this
step depends on the form of the score.

\emph{Step 3: the weighted score distribution.}  For the quantile-regression score,
$V(w,y;\Ztr)=\hat q(\alpha\mid w)-y$.  Define, for $t\in\R$,
\begin{align*}
 H(t)=\E_{P'}\bigl[a(W)\,\P(\hat q(\alpha\mid W)-Y\leq t\mid W,\Ztr)\ \big|\ \Ztr\bigr],
\end{align*}
the distribution function of the calibration score under the weighted distribution.  Since
$Y=\hat s(X)\wedge c_0$, one has $\P(Y\geq v\mid W=w,\Ztr)=\bar F(v\mid w)$ for
$v\leq c_0$.  Assumption~\ref{as:D} at $t=0$ gives $\bar F(q(w)\mid w)=1-\alpha$, and
together with the monotonicity of $\bar F(\cdot\mid w)$, for $0<t\leq r$,
\begin{align}
 \bar F(q(w)+t\mid w)\leq1-\alpha-b_1t,\qquad
 \bar F(q(w)-t\mid w)\geq1-\alpha+b_1t.
 \label{crossbounds}
\end{align}
Fix $h\in(0,r)$.  On the set $\{e(w)\leq h/2\}$ we have
$\hat q(\alpha\mid w)-h\leq q(w)-h/2\leq c_0$ and $\hat q(\alpha\mid w)+h\geq q(w)+h/2$,
with $q(w)+h/2<c_0$, so by \eqref{crossbounds}
\begin{align*}
 \P(\hat q(\alpha\mid w)-Y\leq h\mid W=w,\Ztr)
 &=\P(Y\geq\hat q(\alpha\mid w)-h\mid W=w,\Ztr)\\
 &\geq\bar F(q(w)-h/2\mid w)\geq1-\alpha+b_1h/2,\\
 \P(\hat q(\alpha\mid w)-Y\leq-h\mid W=w,\Ztr)
 &=\P(Y\geq\hat q(\alpha\mid w)+h\mid W=w,\Ztr)\\
 &\leq\bar F(q(w)+h/2\mid w)\leq1-\alpha-b_1h/2.
\end{align*}
Let $D_h=\E_{P'}[a(W)\ind\{e(W)>h/2\}\mid\Ztr]$.  Since the conditional probabilities
lie in $[0,1]$, $\E_{P'}[a(W)\mid\Ztr]=1$ and $1-\alpha+b_1h/2\leq1$, the latter by the second inequality in
\eqref{crossbounds} and $\bar F\leq1$, we get
\begin{align}
 H(h)&\geq(1-\alpha+b_1h/2)(1-D_h)\geq1-\alpha+b_1h/2-D_h,\nonumber\\
 H(-h)&\leq(1-\alpha-b_1h/2)(1-D_h)+D_h\leq1-\alpha-b_1h/2+D_h.
 \label{scorecrossing}
\end{align}
Since $\ind\{e>h/2\}\leq2e/h$, $a\leq\hat \pi^{-1}$ and $\pi\leq1$, we may further bound
\begin{align*}
 D_h\leq\frac2h\,\E_{P'}\Bigl[\frac{e(W)}{\hat \pi(W)}\ \Big|\ \Ztr\Bigr]
 =\frac{2}{h\bar\pi}\,\E_P\Bigl[\frac{e(W)\pi(W)}{\hat \pi(W)}\ \Big|\ \Ztr\Bigr]
 \leq\frac{2}{h\pi_0}\,\E_P\Bigl[\frac{e(W)}{\hat \pi(W)}\ \Big|\ \Ztr\Bigr],
\end{align*}
whose expectation over $\Ztr$ tends to zero by Assumption~\ref{as:E}; hence $D_h\to0$ in
probability.

\emph{Step 4: weighted laws of large numbers.}  Given $\Ztr$ and the retention
indicators, the retained pairs are i.i.d.\ with covariate distribution $P'$.  For fixed $t\in\R$
put $\xi_i=a(W_i)\ind\{V_i\leq t\}$, $i\in\Icap$; these are i.i.d.\ with mean $H(t)$ and
$\E[\xi_i^{1+\rho}]\leq\E_{P'}[a(W)^{1+\rho}\mid\Ztr]\leq M_0:=M/\pi_0$ by
\eqref{weightmoments}.  For $K>0$ write $\xi_i=\xi_i\ind\{\xi_i\leq K\}+\xi_i\ind\{\xi_i>K\}$.
The truncated summands have variance at most $K^2$, and
$\E[\xi_i\ind\{\xi_i>K\}]\leq\E[\xi_i^{1+\rho}]K^{-\rho}\leq M_0K^{-\rho}$.  Hence, for
$\varepsilon\in(0,1)$, by Chebyshev's and Markov's inequalities,
\begin{align*}
 \P\Bigl(\Bigl|\frac1m\sum_{i\in\Icap}\xi_i-H(t)\Bigr|>3\varepsilon\ \Big|\ m\Bigr)
 \leq\frac{K^2}{m\varepsilon^2}+\frac{M_0K^{-\rho}}{\varepsilon}
 +\ind\{M_0K^{-\rho}>\varepsilon\}.
\end{align*}
For fixed $\varepsilon$, choose $K$ so that the indicator term vanishes and the
tail term is arbitrarily small.  Keeping this $K$ fixed, let $m\to\infty$ to make the
variance term vanish.  The bounds depend only on $(M_0,\rho)$, so the convergence is
uniform in the training fold.  Since Step~1 shows that $m\to\infty$ in probability
uniformly in the training fold, the same convergence holds for the random retained
sample size.

The same argument with $\ind\{V_i\leq t\}$ replaced by $1$ gives denominator convergence, thus giving overall
\begin{align*}
 m^{-1}\sum_{i\in\Icap}a(W_i)\ind\{V_i\leq t\}-H(t)&\overset{\P}\longrightarrow0,\quad m^{-1}\sum_{i\in\Icap}a(W_i)\overset{\P}\longrightarrow1.
\end{align*}
It remains to control the weight of a new observation.  By \eqref{weightmoments},
$\E_P[a(W)\mid\Ztr]\leq(\E_P[a(W)^{1+\rho}\mid\Ztr])^{1/(1+\rho)}\leq M^{1/(1+\rho)}$,
so $a(W)=O_{\P}(1)$ and $a(W)/m\to0$ in probability.  Thus, for finite
$t$,
\begin{align*}
 Q_W([-\infty,t])=
 \frac{m^{-1}\sum_{i\in\Icap}a(W_i)\ind\{V_i\leq t\}}
      {m^{-1}\sum_{i\in\Icap}a(W_i)+a(W)/m}
 =H(t)+o_{\P}(1).
\end{align*}
\emph{Step 5: the cutoff.}  By \eqref{scorecrossing} and Step~4, with probability
tending to one,
\begin{align*}
 Q_W([-\infty,h])&\geq1-\alpha+b_1h/2-D_h-o_{\P}(1)>1-\alpha,\\
 Q_W([-\infty,-h])&\leq1-\alpha-b_1h/2+D_h+o_{\P}(1)<1-\alpha.
\end{align*}
On this event,
$z\mapsto Q_W([-\infty,z])$ being nondecreasing, the definition of the quantile gives
$\eta(W)\leq h$, and $Q_W([-\infty,z])<1-\alpha$ for every $z\leq-h$ gives
$\eta(W)\geq-h$.  As $h\in(0,r)$ is arbitrary, $\eta(W)\to0$ in probability.

Since
$\hat \pi\leq1$, $\E[e(W)]\leq\E[e(W)/\hat \pi(W)]\to0$ by Assumption~\ref{as:E}, so
$e(W)\to0$ in probability and so we may write $\hat q(\alpha\mid W)=q(W)+o_{\P}(1)$.  Hence
\begin{align*}
 \hat q(\alpha\mid W)-\eta(W)=q(W)+o_{\P}(1).
\end{align*}

Moreover, since $q(W)+r<c_0$, the untruncated bound is smaller
than $c_0$ with probability tending to one.  Hence, on that event, \eqref{CQRbound} gives
\begin{align*}
 \widehat L(W)=\hat q(\alpha\mid W)-\eta(W)=q(W)+o_{\P}(1),
\end{align*}
which proves \eqref{qconsistency}.

\emph{Step 6: coverage.}  Given $W$ and the data, $\widehat L(W)$ is fixed and the
 vector $X$ of the new observation is independent of the data, so
\begin{align*}
 \P(X\in\widehat R(W)\mid W,\Ztr,\Zca)
 =\P(\hat s(X)\geq\widehat L(W)\mid W,\Ztr,\Zca)=\bar F(\widehat L(W)\mid W),
\end{align*}
since $X$ is independent of $\Zca$ given $(W,\Ztr)$.  On the event
$\{|\widehat L(W)-q(W)|\leq r\}$, Assumption~\ref{as:D} gives
$|\bar F(\widehat L(W)\mid W)-(1-\alpha)|\leq b_2|\widehat L(W)-q(W)|$, which tends to
zero in probability by \eqref{qconsistency}; this is \eqref{condcoverage}.  The
conditional coverage lies in $[0,1]$, so convergence in probability also gives
convergence of its expectation, which is
$\P(X\in\widehat R(W))=\E[\bar F(\widehat L(W)\mid W)]\to1-\alpha$.
\end{proof}

\subsection*{Proofs for Section~\ref{sec:price}}

\begin{proof}[Proof of Proposition~\ref{prop:retention}]
For comonotone coordinates the survival copula is
$\mathcal K_C(u)=\min_ju_j$ and for independent coordinates it is
$\mathcal K_C(u)=\prod_ju_j$ \citep[Ch.~2]{Nelsen2006}.  With equal margins $\bar G_0$,
\eqref{piformula} gives $\bar G_\psi(c)=\bar G_0(c)$ in the comonotone case and
$\bar G_\psi(c)=\bar G_0(c)^d$ in the independent case, for every threshold $c$ in the
range.  Hence the set $\{c:\bar G_\psi(c)\geq\pi_0\}$ is $\{c:\bar G_0(c)\geq\pi_0\}$
in the comonotone case and is $\{c:\bar G_0(c)\geq\pi_0^{1/d}\}$ in the independent
case; taking suprema gives \eqref{retentionfloor} by the definition \eqref{ginv}.

For $\bar G_0(c)=e^{-\gamma c}$
the equations $e^{-\gamma c}=\pi_0$ and $e^{-\gamma c}=\pi_0^{1/d}$ have the solutions
$-\log\pi_0/\gamma$ and $-\log\pi_0/(d\gamma)$, respectively.
\end{proof}

\begin{proof}[Proof of Proposition~\ref{prop:naive}]
(i)  The cutoff $\widehat L$ is a function of the sample, and the new observation is
independent of the sample, so conditioning on the sample gives
$\P(\rZ\geq\widehat L)=\E[\bar H_\psi(\widehat L)]$, hence
$\E[\bar H_\psi(\widehat L)]\geq1-\alpha$.  On $A=\{\widehat L>v^H_\psi+\varepsilon\}$,
monotonicity of $\bar H_\psi$ gives $\bar H_\psi(\widehat L)\leq p_\varepsilon$, and
elsewhere $\bar H_\psi(\widehat L)\leq1$.  Hence
\begin{align*}
 1-\alpha\leq\E[\bar H_\psi(\widehat L)]
 \leq p_\varepsilon\P(A)+1-\P(A).
\end{align*}
Rearranging gives $\P(A)(1-p_\varepsilon)\leq\alpha$.

If $v>\mathop{\rm ess\,sup}\rC$, then $\rZ\leq\rC<v$ almost surely, so
$\{\rZ\geq\widehat L\}\subseteq\{\widehat L<v\}$ up to a null set, and
$1-\alpha\leq\P(\rZ\geq\widehat L)\leq\P(\widehat L<v)$.

(ii)  The function $v\mapsto\bar H_\psi(v)=\P(\rZ\geq v)$ is left continuous, since
$\{\rZ\geq v_k\}\downarrow\{\rZ\geq v\}$ as $v_k\uparrow v$.  By the definition of
$v^H_\psi$ there are $v_k\uparrow v^H_\psi$ with $\bar H_\psi(v_k)\geq1-\alpha$, so
$\bar H_\psi(v^H_\psi)\geq1-\alpha$.  Under Assumption~\ref{as:coic}, \eqref{ZisXC}
gives $\bar F_\psi(v^H_\psi)\bar G_\psi(v^H_\psi)\geq1-\alpha>0$; in particular
$\bar G_\psi(v^H_\psi)>0$, and division gives the bound.

If $\bar H_\psi$ is continuous
at $v^H_\psi$ then also $\bar H_\psi(v^H_\psi)\leq1-\alpha$, because
$\bar H_\psi(v^H_\psi+1/k)<1-\alpha$ for every $k$ by the definition of the supremum, so
equality holds throughout.
\end{proof}

\begin{proof}[Proof of Proposition~\ref{prop:bonf}]
Continuity of $\bar F_{\psi,1}$ and finiteness of the levels give
$\bar F_{\psi,j}(v^{\rm B}_\psi)=1-\alpha/d$ for every $j$, where $v^{\rm B}_\psi$ is the
common value of the $v^{\rm B}_{\psi,j}$, so $R^{\rm B}_\psi=R_\psi(v^{\rm B}_\psi)$.  The
union bound gives
\begin{align}
 \bar F_\psi(v^{\rm B}_\psi)
 =1-\P\Bigl(\bigcup_{j=1}^d\{\psi_j(X^{(j)})<v^{\rm B}_\psi\}\Bigr)
 \geq1-\sum_{j=1}^d\P(\psi_j(X^{(j)})<v^{\rm B}_\psi)=1-\alpha.
 \label{unionbd}
\end{align}
Hence $v^F_\psi\geq v^{\rm B}_\psi$ by \eqref{corners}, and since $v\mapsto R_\psi(v)$ is
nonincreasing, \eqref{bonfcontainment} follows.

Equality in \eqref{unionbd} holds
exactly when the probability of the union equals the sum of the probabilities, which is
equivalent to pairwise disjointness of the events up to null sets; this proves the
equivalence.

If the events are not pairwise disjoint then
$\bar F_\psi(v^{\rm B}_\psi)>1-\alpha$.  Each $\psi_j(X^{(j)})$ is atomless, because its
survival function is continuous, so $\P(\rX=t)\leq\sum_j\P(\psi_j(X^{(j)})=t)=0$
and $\bar F_\psi$ is continuous.  Therefore $\bar F_\psi(v)\geq1-\alpha$ for some
$v>v^{\rm B}_\psi$, and so by definition $v^F_\psi>v^{\rm B}_\psi$.

\end{proof}

\subsection*{Proofs for Section~\ref{sec:design}}

\begin{proof}[Proof of Lemma~\ref{lem:maximin}]
Let $u\in(0,1)^d$ satisfy $\mathcal K_X(u)=1-\alpha$ and put $m=\min_j(-\log u_j)$, so
that $u_j\leq e^{-m}$ for every $j$.  A copula is nondecreasing in each argument, so
\begin{align*}
 1-\alpha=\mathcal K_X(u)\leq\mathcal K_X(e^{-m},\ldots,e^{-m})=\Phi(e^{-m}).
\end{align*}
Since $\Phi(u)<1-\alpha$ for $u<u^*$, this forces $e^{-m}\geq u^*$, that is,
$m\leq-\log u^*$.  The diagonal vector $(u^*,\ldots,u^*)$ satisfies the constraint,
because $\Phi(u^*)=1-\alpha$, and takes the value $-\log u^*$.
\end{proof}

\begin{proof}[Proof of Theorem~\ref{thm:pathstability}]
Work on $\{\hat\psi\in\mathcal N(\varepsilon_N)\}$ and take $N$ large enough that
$\varepsilon_N<\delta$.  For each $j$, choose $t_j$ such that
$\psi_j(t_j)=c_0+\delta$.  We claim that, for every $v\leq c_0$,
\begin{align*}
 \{x:\psi_j(x)\geq v+\varepsilon_N\}
 \subseteq\{x:\hat\psi_j(x)\geq v\}
 \subseteq\{x:\psi_j(x)\geq v-\varepsilon_N\}.
\end{align*}
On the set where $\psi_j(x)\leq c_0+\delta$, both inclusions follow from the
bound $\psi_j(x)-\varepsilon_N\leq\hat\psi_j(x)\leq
\psi_j(x)+\varepsilon_N$.  If $\psi_j(x)>c_0+\delta$, then $x>t_j$ and
monotonicity gives
\begin{align*}
 \hat\psi_j(x)\geq\hat\psi_j(t_j)\geq c_0+\delta-\varepsilon_N>c_0.
\end{align*}
This proves the first inclusion in the remaining case; the second is immediate from
$\psi_j(x)>c_0+\delta>v-\varepsilon_N$.  Intersecting over $j$ proves
\eqref{pathsandwich}.  Applying the same inclusions to $C$ at $v=c_0$ proves
\eqref{pathretention}, while conditional probabilities for $X$ give
\begin{align*}
 \bar F_\psi(v+\varepsilon_N\mid w)
 \leq\bar F_{\hat\psi}(v\mid w)
 \leq\bar F_\psi(v-\varepsilon_N\mid w),\qquad v\leq c_0.
\end{align*}

For a nonincreasing function $S$, if
$q=\sup\{v:S(v)\geq1-\alpha\}$, then $S(v)\geq1-\alpha$ for $v<q$ and
$S(v)<1-\alpha$ for $v>q$.  The truncation in $q_\psi(w)$ is inactive because
$q_\psi(w)<c_0$, so this observation applies with
$S(v)=\bar F_\psi(v\mid w)$.  We may also take $N$ large enough that
$q_\psi(w)+2\varepsilon_N<c_0$ for every $w$.  The probability double inequality shows that
$\bar F_{\hat\psi}(v\mid w)\geq1-\alpha$ when
$v<q_\psi(w)-\varepsilon_N$, and that $\bar F_{\hat\psi}(v\mid w)<1-\alpha$ when
$q_\psi(w)+\varepsilon_N<v\leq c_0$.  Since $q_{\hat\psi}(w)\leq c_0$ by
definition, it follows that
\begin{align*}
 q_\psi(w)-\varepsilon_N
 \leq q_{\hat\psi}(w)
 \leq q_\psi(w)+\varepsilon_N,
\end{align*}
uniformly in $w$.  This proves \eqref{pathquantile}.

The monotonicity of the regions and \eqref{pathsandwich} now yield
\begin{align*}
 R_\psi(q_\psi(w)+2\varepsilon_N)
 &\subseteq R_{\hat\psi}(q_\psi(w)+\varepsilon_N)
 \subseteq R_{\hat\psi}(q_{\hat\psi}(w))\\
 &\subseteq R_{\hat\psi}(q_\psi(w)-\varepsilon_N)
 \subseteq R_\psi(q_\psi(w)-2\varepsilon_N),
\end{align*}
and hence \eqref{pathregion}.  Taking conditional probabilities of the outer sets gives
\begin{align*}
 \bar F_\psi(q_\psi(w)+2\varepsilon_N\mid w)
 &\leq \P(X\in R_{\hat\psi}(q_{\hat\psi}(w))\mid W=w,\Ztr)\\
 &\leq \bar F_\psi(q_\psi(w)-2\varepsilon_N\mid w).
\end{align*}
Once $2\varepsilon_N\leq r$, Assumption~\ref{as:D} bounds both endpoints within
$2b_2\varepsilon_N$ of $1-\alpha$, uniformly in $w$.  This proves \eqref{pathcoverage}.
\end{proof}

\end{document}